\documentclass[9pt,a4paper,twocolumn,twoside]{rho-class/rho}
\usepackage[english]{babel}
\usepackage{graphicx}
\usepackage{subcaption}

\title{A Structural Comparison of Indian Epics}

\author[a]{J.D.P. Evans, University of Hertfordshire, \href{mailto:j.evans8@herts.ac.uk}{j.evans8@herts.ac.uk}}
\journalname{arXiv}
\theday{\today}
\begin{abstract}
    The \textit{Mah\={a}bh\={a}rata} and {R\={a}mayana constitute the two major Sanskrit epics of ancient India. Traditionally they are viewed quite differently, with the former being seen as more historical and the latter more poetic. We perform network analysis on both texts in an attempt to gather quantitative information on interrelationships between characters, thereby offering quantitative insight into such separate classifications. We also compare with the \textit{Iliad} in an attempt to see if structural similarities exist that underlie several narrative similarities.}
\end{abstract}

\keywords{Network Analysis, Narrative Networks, Complex Systems, Indian Epics, Social Networks}

\begin{document}

    \maketitle
    \thispagestyle{firststyle}

%----------------------------------------------------------

\section{Introduction}
\label{sec:Introduction}

In the field of comparative mythology, complex networks have been used to offer quantitative analysis of epic literature, cf. \cite{JanickyjMacCarronKenna2024}, \cite{MacCarronKenna2013}, \cite{MacCarronKenna2012}, \cite{YoseMacCarron2018}, \cite{YoseMacCarron2016}. Such insights have their genesis in statistical physics and complexity theory where classification of critical phenomena into universality classes is natural. Different classes of network have been identified according to various properties (cf. \cite{Amaral2000}, \cite{WattsStrogatz1998}) with a notable class being that of real-world social networks. These tend to share distinguishing properties commonly associated with them, see Section \ref{sec:universal_properties}. Using such observations, an early application of network analysis to epic literature was in \cite{MacCarronKenna2012}, in which three iconic European epic narratives were compared, not only to each other but also to real, imaginary, and random networks. These epic narratives were shown to exhibit some or all of the properties typically associated with real social networks, at least to some degree.

While initial interest in comparisons with real-world social networks has diminished, the techniques developed in \cite{MacCarronKenna2012} nevertheless highlights the benefits of quantitative comparisons, even going so far as attempted classifications of epic literature. Thus far, emphasis has largely been on European texts, such as those of Ireland (see \cite{MacCarronKenna2012}, \cite{YoseMacCarron2018}), Iceland (see \cite{MacCarronKenna2013}), Scotland (see \cite{YoseMacCarron2016}), Greece (see \cite{MacCarronKenna2012}). Biblical networks have also been examined (see \cite{Massey2016}). However, comparatively little has been done for the Indian epics. In \cite{JanickyjMacCarronKenna2024} there is a brief mention of the \textit{Mah\={a}bh\={a}rata} in the form of a comparison of female representation across mythologies, and in \cite{GultepeMathangi2023} there is a social network analysis of the character relationships in the \textit{Mah\={a}bh\={a}rata}. As we shall see in Section \ref{sec:network_construction}, the methodology employed in \cite{GultepeMathangi2023} is fundamentally different to those employed in this monograph however. In \cite{GultepeMathangi2023}, their network was generated using LSA word vectors, whereas we employ a more straightforward if time-consuming approach in which characters are represented by nodes and edges represent character interactions. The approach of \cite{GultepeMathangi2023} was to essentially construct edges if two characters appear in semantically similar textual environments, whereas our approach is far more explicit. The benefit of doing so, we argue, is that it greatly improves interpretability and validity when posing questions of comparative mythology.

With this in mind, we shall undertake two key comparisons. First, we compare the \textit{Mah\={a}bh\={a}rata} and the \textit{R\={a}m\={a}ya\d{n}a} at the structural level. As we shall see, the texts differ in several key ways. Such differences may reflect the cultural differences between these two texts, with the \textit{Mah\={a}bh\={a}rata} being viewed in India in a much more historical light than the \textit{R\={a}m\={a}ya\d{n}a}, which is often viewed more as `the first poetic work'. However, we argue that the observed differences are not sufficient to suggest such distinct classification. Second, we shall make comparisons with the great Western epic, the \textit{Iliad}. This has many narrative similarities with the \textit{Mah\={a}bh\={a}rata} in particular, as noted in \cite{SmithMaha}. As such, we offer a quantitative insight into this comparison. We shall also be comparing these networks to random networks, thereby providing unique insights into these important texts in Indian cultural heritage.

In Section \ref{sec:network_construction}, we shall explain the construction of these networks. In Section \ref{sec:epic_sanskrit_texts}, we shall present a brief overview of the texts to contextualise what follows, while in Sections \ref{sec:universal_properties}-\ref{sec:relationship_with_western_texts} we shall present the analysis itself. We conclude in Section \ref{sec:conclusion}. A brief summary of Sanskrit pronunciation can be found in Section \ref{app:pronouncing_sanskrit}.

\section{Network Construction}
\label{sec:network_construction}

Construction of the respective networks began with a thorough reading of the texts. Each character is represented as a node, alongside other information including the character's gender. Edges are then constructed when two characters interact physically in some way, or are related either by blood or marriage. Physical interactions can include speaking to each other directly or being part of a group in which it is clear that the two characters know each other. This allows for situations in which a group of characters are present but only a strict subset speak in the text, and avoids situations such as two characters both being present at an event (such as a celebration) but in which they do not actually meet or interact. Additionally, we differentiate between hostile and friendly interactions\footnote{We shall use interchangeably the terms positive/negative and friendly/hostile, respectively.}. We define a hostile interaction to be any interaction in which there is some form of violence. Primarily, this is physical but could also include other forms of violence such as a threat. A friendly interaction is then defined to be any other type of interaction. Note that all interactions, friendly or hostile, are symmetric and so our networks will necessarily be undirected.

\begin{figure*}[t]
    \centering

    \begin{subfigure}[b]{0.32\textwidth}
        \centering
        \includegraphics[width=\textwidth]{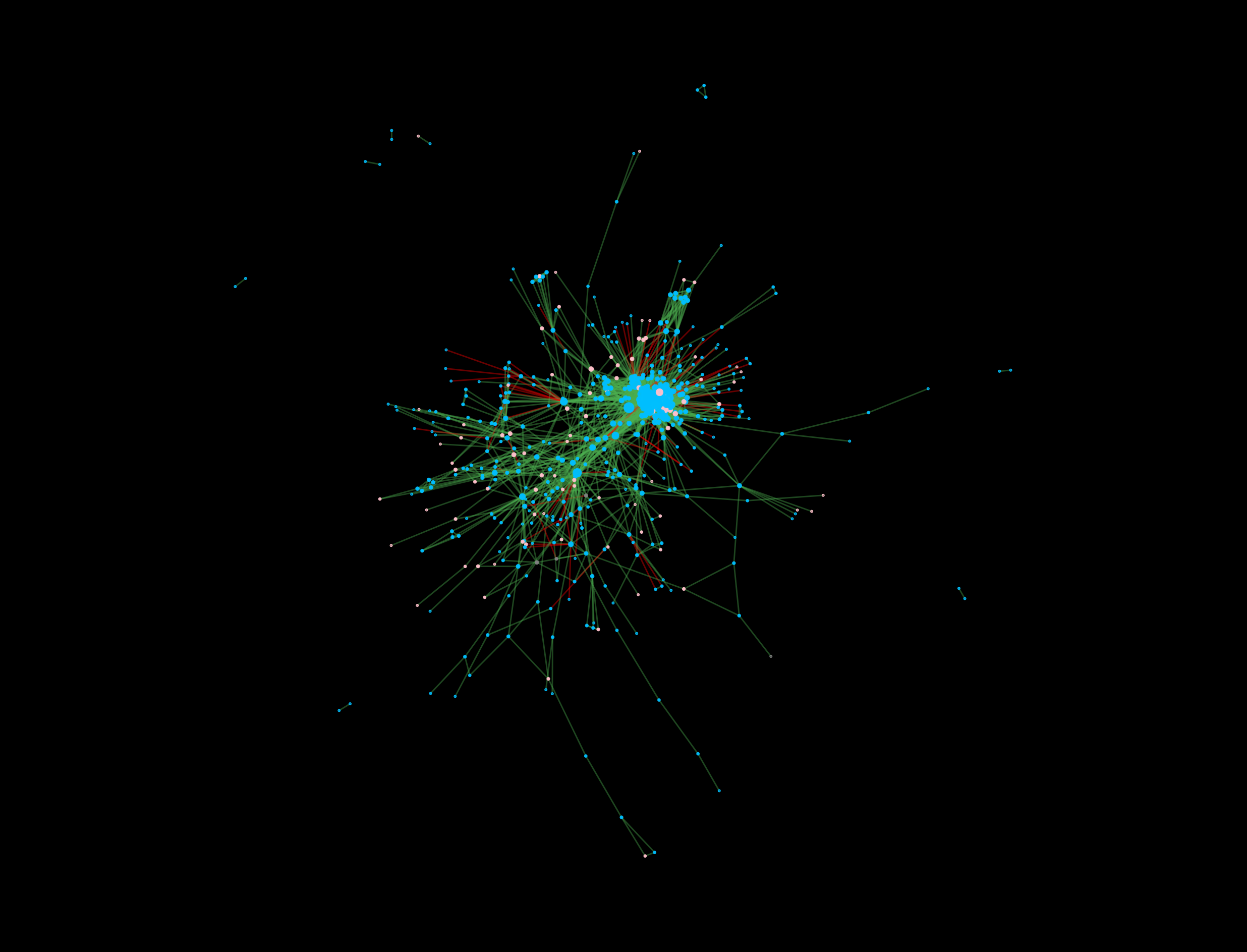}
        \caption{\textit{Mah\={a}bh\={a}rata}}
    \end{subfigure}
    \hfill
    \begin{subfigure}[b]{0.29\textwidth}
        \centering
        \includegraphics[width=\textwidth]{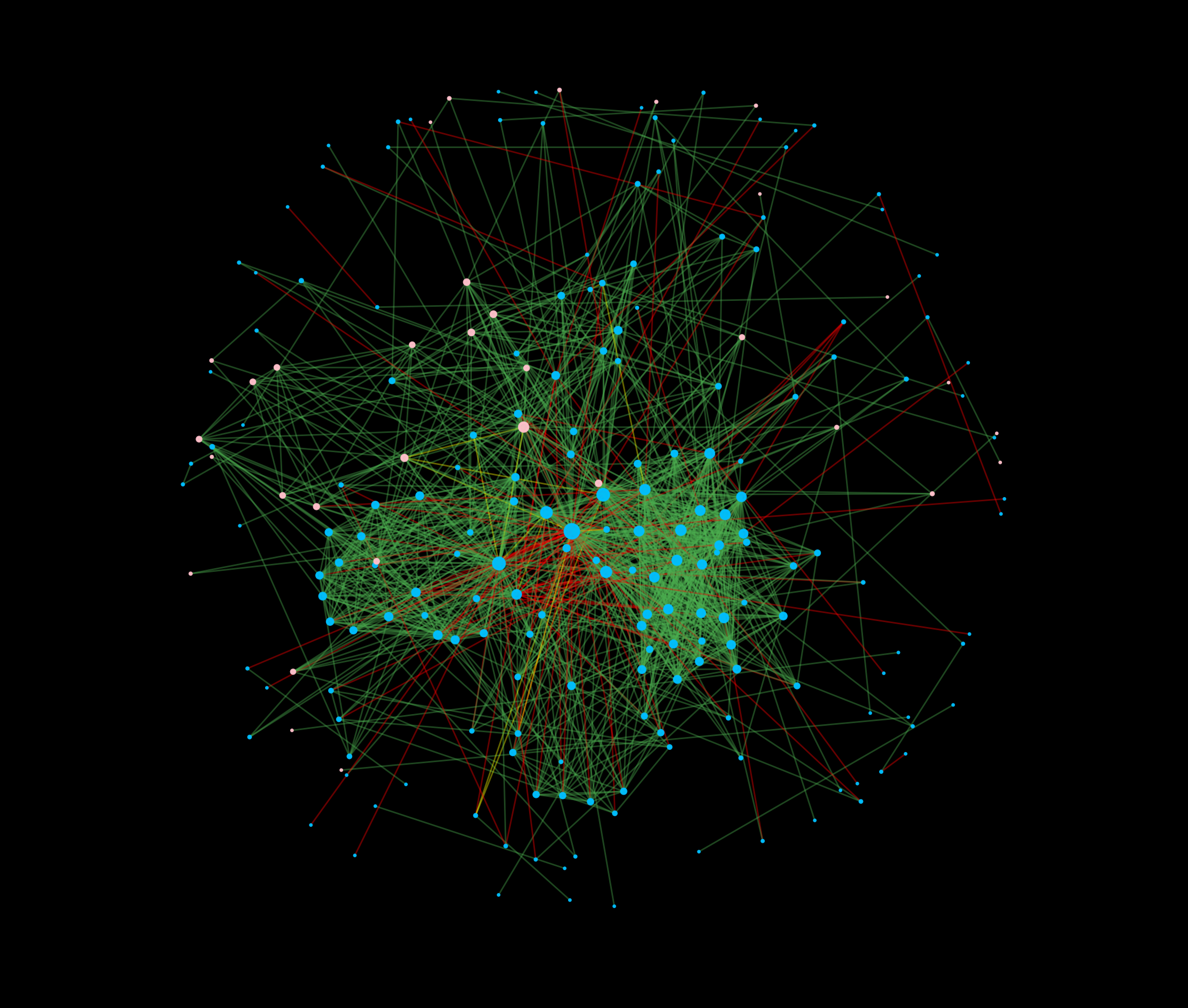}
        \caption{\textit{R\={a}m\={a}ya\d{n}a}}
    \end{subfigure}
    \hfill
    \begin{subfigure}[b]{0.285\textwidth}
        \centering
        \includegraphics[width=\textwidth]{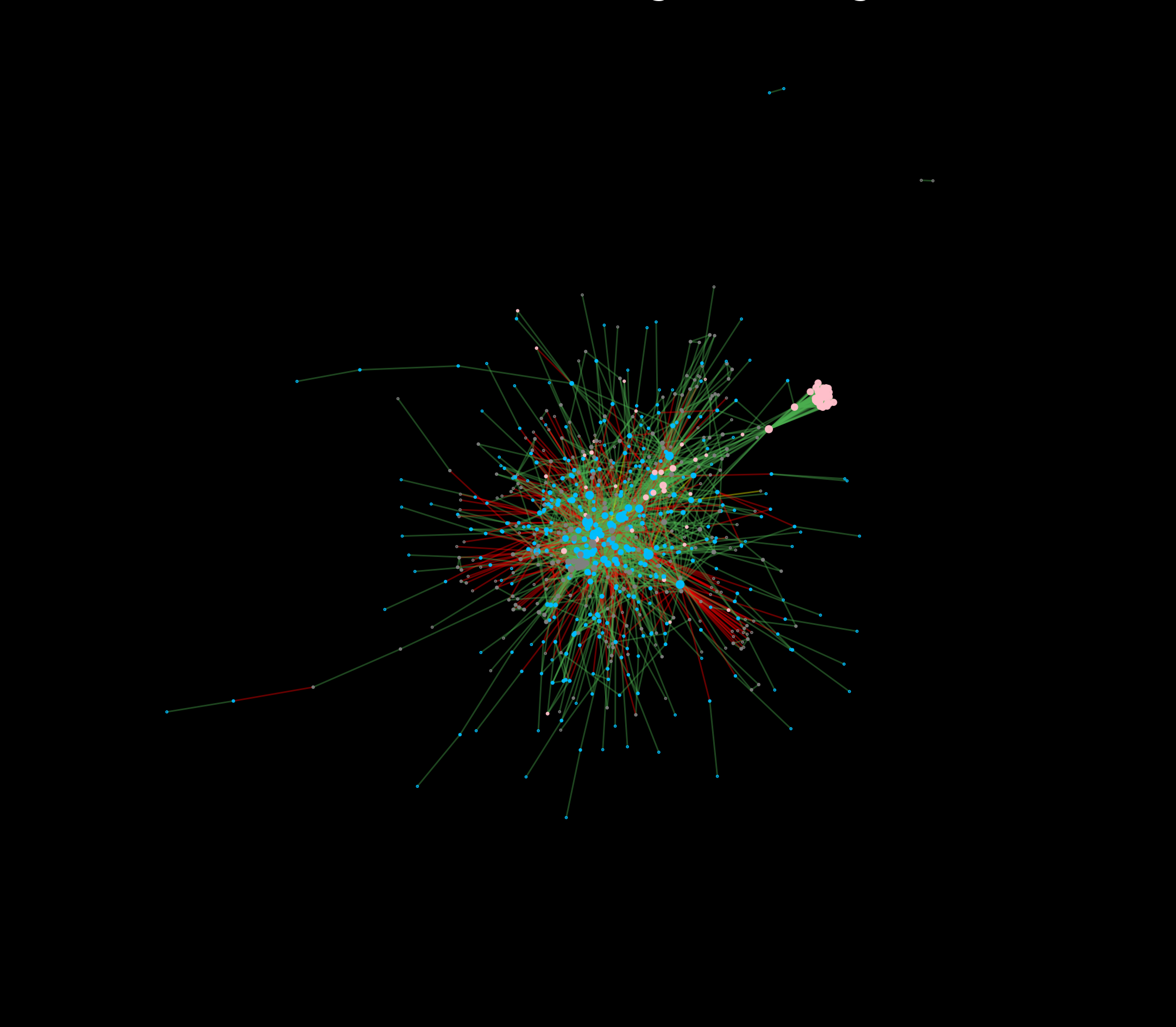}
        \caption{\textit{Iliad}}
    \end{subfigure}

    \caption{Three main networks. Blue nodes represent male characters, pink nodes represent female characters. Green edges represent friendly edges, red edges represent hostile.}
    \label{fig:networks}
\end{figure*}

Following the above definitions, we obtain the entire character networks for the \textit{Mah\={a}bh\={a}rata}, \textit{R\={a}m\={a}ya\d{n}a} and \textit{Iliad}. Each of these are represented in \figref{fig:networks} in which we represent male (resp. female) nodes by blue (resp. pink) nodes. We represent friendly (resp. hostile) edges by green (resp. red) edges. Note that it is possible for there to be both friendly and hostile interactions. Such mixed edges are rare and are represented by the colour yellow.

\section{Epic Sanskrit Texts}
\label{sec:epic_sanskrit_texts}

The \textit{Mah\={a}bh\={a}rata} is an Indian epic, written sometime between 400BCE-400CE\footnote{Note that ancient Indian chronology is notoriously difficult to establish and the \textit{Mah\={a}bh\={a}rata} is no exception.}. Traditionally, it is said to contain 100,000 verses with most verses consisting of two lines each. A critical shortened edition is the Poona edition (published between 1933 and 1966), which retains 67,314 verses and is still considered to be authentic. For this work, we use the heavily abridged translation by J.D. Smith (see \cite{SmithMaha}) which is based upon the Poona edition. Given the heavily condensed nature of Smith's translation, all analysis which follows should keep this in mind. To give a sense of scale, in S. S\"{o}rensen's \textit{An Index to the Names in the Mahabharata} \cite{Sorensen1978}, the list runs to 807 two-column pages. In contrast, our network has just 537 characters.

The reason we use Smith's translation\footnote{Note that Smith estimates he translated just under 11\% of the entire text.} is two-fold. First, its size is manageable. As the above suggests, to work with the entire text (or even the Poona edition) would be a much larger undertaking, likely resulting in an extraordinary data collection time period. Given the relative youth of this area of research, this seems counterproductive. Rather, a full examination of the entire text seems more suitable for a future research direction, ideally utilising a team of researchers. The second reason is that Smith's aim was to produce an English version of the \textit{Mah\={a}bh\={a}rata} that is `as true to the original' as possible. We argue that Smith's attempt captures the broad structural properties of the full text. As Smith notes, the full text is highly circuitous including `digressions within digressions', repetition and large amounts of theological doctrine. We do not argue that such material is inconsequential, only that its removal is unlikely to impact the overall structure to such an extent that our intended comparisons are invalid. A final point on the translation used in this study is that of style. The \textit{Mah\={a}bh\={a}rata} is composed almost entirely in verse, whereas Smith's translation is into English prose. Smith's justification for this is to encourage readability of the text. As we are interested solely in characters and their interactions, we argue such an alteration will have limited impact on the final results. Indeed, the literal translation is far more likely to have an impact on this than the move to English prose. For this, we must trust the translator and in this regard Smith is highly respected.

At its most basic, the \textit{Mah\={a}bh\={a}rata} is a story about a conflict between two sets of cousins. It includes the events which led to war, the eighteen-day war itself, and the aftermath of this war. However, it is also a story which espouses the concept of \textit{dharma}. Roughly, this is the notion of doing what is right for a person to do. As such, what is right for one person, may be wrong for another. Understanding what \textit{dharma} is appropriate for a person is key to the attainment of good \textit{karma}, which is itself key to the cycle of rebirth. A typical thought experiment proceeds thus: if violence is wrong, and leads to bad \textit{karma}, thus leading to more rebirths, how is the warrior and kingly caste (\textit{K\d{s}atriya}) to behave? This theme is most famously addressed in the \textit{Bhagavadg\={i}t\={a}}, the sermon preached by K\d{r}\d{s}\d{n}a to Arjuna on the eve of the great war. \textit{Dharma} is the \textit{Mah\={a}bh\={a}rata}'s attempt to address such questions. Understanding one's \textit{dharma} comes down to caste, stage of life and even circumstance. For instance, at 12.283, it is stated that a Brahmin in need may follow the \textit{dharma} of a K\d{s}atriya or Vai\'{s}ya, \cite{SmithMaha}. It is a theme throughout the \textit{Mah\={a}bh\={a}rata}.

It is therefore important to stress that our methods cannot deal with questions relating to this deeper meaning behind the \textit{Mah\={a}bh\={a}rata}.  What they can achieve however, is a structural comparison of such epics with, for example, Western counterparts. For as noted by Smith in \cite{SmithMaha}, `The story told in the \textit{Mah\={a}bh\={a}rata} does bear similarities, both large-scale and small-scale, to other epic narratives. The style of the text is reminiscent of other works thought to have originated as orally composed bardic epics.' A key point of comparison will therefore be with the \textit{Iliad}, the Greek epic attributed to Homer and dated to the Eighth Century BCE, \cite{Iliad2003}.

Another point of comparison, and one which shall take up the majority of this monograph, is that of the parallel tale of R\={a}ma. We shall use an early form of the \textit{R\={a}m\={a}ya\d{n}a} \cite{BrockingtonRama} translated by J. Brockington and M. Brockington called \textit{R\={a}m\={a} the Steadfast}. Likely to have originated as an oral composition (cf. \cite{BrockingtonRama}), the date of its composition is still a matter of debate. J. \& M. Brockington suggest a date of roughly the Fifth Century BCE\footnote{This follows from linguistic evidence based upon texts which can be roughly dated. A date of First Century CE is reasonable for the written date.}, although many additions and changes took place over the following centuries, as was the case for the \textit{Mah\={a}bh\={a}rata}. Our translation comes from the earliest levels of the story and represents an early stage of its development. The reason for this earlier form is again two-fold. First, as with the \textit{Mah\={a}bh\={a}rata}, it offers a digestible account of the text that we argue provides a meaningful insight into the structure of the whole text. Second, while the \textit{R\={a}m\={a}ya\d{n}a} is justifiably called an `epic', grand in scope much as is the \textit{Mah\={a}bh\={a}rata}, the label of `epic' is often seen as more suitable to the later versions of the \textit{R\={a}m\={a}ya\d{n}a}. As such, this provides another point of comparison. 

We should also note different origins of the two texts. In the Introduction to \textit{R\={a}m\={a} the Steadfast}, the translators compare it to a Western heroic romance. In it, we meet the hero R\={a}ma, now deified following centuries of Indian tradition. Forced into exile by his father at the behest of his father's wife (not R\={a}ma's mother), he is joined by his brother Lak\d{s}ma\d{n}a and wife S\={i}t\={a}. During their exile, S\={i}t\={a} is kidnapped by the evil R\={a}va\d{n}a and the remainder of the text follows the ensuing war to rescue her. In contrast to the epic scope of the \textit{Mah\={a}bh\={a}rata}, the struggle here is that for integrity and happiness for R\={a}ma. In contrast to Arjuna's famous grappling with what action is permissible for a hero to follow while maintaining integrity, R\={a}ma's struggle is intensely personal. He wishes to fulfill his duty as a son by accepting his father's banishment, and then he wishes to rescue his wife. To once more quote the aforementioned Introduction, ``Indian tradition too distinguishes between the two works, designating the \textit{R\={a}m\={a}ya\d{n}a} as the \textit{\={a}dik\={a}vya}, `the first poetic work' or perhaps `the first work of pure literature', but the \textit{Mah\={a}bh\={a}rata} commonly as \textit{itih\={a}sa}, `thus indeed it was', a term roughly equivalent to history'." It is therefore our initial goal to discuss what structural differences and similarities exist between these two texts.

As a final comment, we note that later versions of the \textit{R\={a}m\={a}ya\d{n}a} greatly expanded upon the original composition. The Critical Edition contains almost 20,000 verses\footnote{Less than a third of the \textit{Mah\={a}bh\={a}rata} but still nevertheless substantial.} and the hero R\={a}ma is repeatedly reworked, first from a moral hero to a regal one, then as an earthly manifestation (\textit{avt\={a}ra)} of the god Vi\d{s}\d{n}u. Eventually, he became God in his own right, the text thereby acquiring the status of holy scripture. Comparisons with such later editions, and indeed comparisons \textit{between} editions, is left to later work. The version upon which this translation is based is the Vadodar\={a} (Westernised as `Baroda') recension\footnote{A recension is a version of a text that has been produced by someone copying the text and making changes to it.}. Its goal in the second half of the Twentieth Century was to identify a version of the R\={a}ma story that is as close as possible to the original, while still containing the large amount of detail added over two millennia. J. and M. Brockington's translation then performed linguistic analysis on the central five books of the `Baroda' text, identifying and including only those passages deemed to preserve the earlier diction. We direct the reader to the text's Introduction for a more thorough discussion of its construction, though we conclude with a caution in the form of the translators' own words: ``Composition was a continuous process taking place over several centuries and involving many tellers, so `earlier', `later', or simply `different' styles can be discerned within each stage, as can narrative anomalies or contradictions symptomatic of multiple, continuous authorship."

\section{Universal Properties}
\label{sec:universal_properties}

We begin our investigation\footnote{For now, we ignore comparisons with the \textit{Iliad}, leaving this to Section \ref{sec:relationship_with_western_texts}.} by exploring various properties which, taken together, tend to indicate a real-world social network. In \cite{MacCarronKenna2012}, it was hoped that by quantitatively computing these properties that it might be possible to place mythological texts on a spectrum from the real to the imaginary. This perception has shifted somewhat in recent years and so now the viewpoint is less about whether mythological texts reflect real-world social networks, and more about attempted quantitative classification of various mythological texts\footnote{Though, as Smith rightly points out in \cite{SmithMaha}: `...Genres are not categories. A category is an impermeable container, and any given item is definitively either in it or out of it.' Rather, we adopt the term classification in its loosest sense.}, or about obtaining historical insights from the culture in which the text was written. Nevertheless, such a viewpoint is still useful. For one, it provides an initial structure to the investigation, offering if nothing else, a comparative device. In the present context, it also allows us to weigh in on the degree of reality inherent in these texts. In India, \textit{itih\={a}sa} (`historical narrative') is a term commonly used for the \textit{Mah\={a}bh\={a}rata} (see Introduction of \cite{SmithMaha}). As Smith points out, ``For believing Hindus the events it [\textit{Mah\={a}bh\={a}rata}] describes are real happenings that took place in India's ancient past." Smith then goes on to say, ``By contrast, the \textit{R\={a}m\={a}yana}, which tells a comparable story in similar metre, language and style, is more normally thought of as a \textit{k\={a}vya}, `literary composition, poem'; indeed, it is considered to be the \textit{\={a}dik\={a}vya}, the first poem."
\medskip

\noindent\textbf{Question:} Is there a sense in which the \textit{Mah\={a}bh\={a}rata} more closely resembles a real-world social network than does the \textit{R\={a}m\={a}yana}?
\medskip

To once more borrow the words of Smith\footnote{See page 1xv of \cite{SmithMaha}.}, we encounter another comparative device: `It is rather striking that two ancient texts with so much in common should traditionally be assigned to different genres, but the reason probably has more to do with the strong perception of V\={a}lm\={i}ki, author of the \textit{R\={a}m\={a}ya\d{n}a}, as the founder of Sanskrit literary tradition than with any sense that the two works are radically different in character.' As such, we raise a further question:
\medskip

\noindent\textbf{Question:} Is there a structural sense in which the two texts belong in different genres?
\medskip

To address these questions, we start with a review of properties which, if taken together, tend to suggest a real-world social network. We stress that such properties are not exclusive to real-world social networks, nor do all such networks necessarily share these properties. Nevertheless, when all are present, this tends to be indicative of real-world social networks. The properties are as follows:
\begin{itemize}
    \item Right-skewed degree distribution, \cite{Amaral2000},  \cite{BarabasiAlbert1999},  \cite{BroidoClauset2019}, \cite{ClausetShaliziNewman2009};
    \item Small worldness ($l\approx l_{rand},\,C>>C_{rand}$), \cite{Amaral2000}, \cite{WattsStrogatz1998};
    \item High clustering coefficient ($C_T>C_n$), \cite{NewmanPark2003};
    \item Non-negative assortativity, \cite{Newman2002};
    \item Structural balance, \cite{SzellLambiotteThurner2010};
    \item Community and hierarchical Structure, \cite{Fortunato2010}, \cite{RavaszBarabasi2003}.
\end{itemize}

To explain the above, we first make some definitions. First, the average path length $l$ is the average number of edges separating two vertices. When computing this value we need to be careful since in general our networks will be disconnected. As such, computing path lengths between disconnected components leads to undefined path lengths. Typically, to get around this, authors compute the average path length on the giant component of the network, i.e. the largest connected component of the network. While suitable, our approach is slightly different in that we take the average shortest path length over all \textit{reachable} pairs of nodes. Specifically, we compute the average shortest path length for each connected component and then take the weighted average in which weight reflects the number of node pairs in that component. Isolated nodes do not contribute.

In our networks the giant component of the \textit{Mah\={a}bh\={a}rata} is made up of 96.65\% of all nodes and 99.62\% of all edges, while the giant component of the \textit{R\={a}m\={a}ya\d{n}a} is made up of 81.78\% of nodes and 99.12\% of edges. The \textit{Mah\={a}bh\={a}rata} therefore has a much larger giant component, relatively speaking, meaning there are fewer isolated nodes. However, in both cases, nearly all edges are in the giant components.

\begin{figure*}[t]
    \centering

    \begin{subfigure}[b]{0.315\textwidth}
        \centering
        \includegraphics[width=\textwidth]{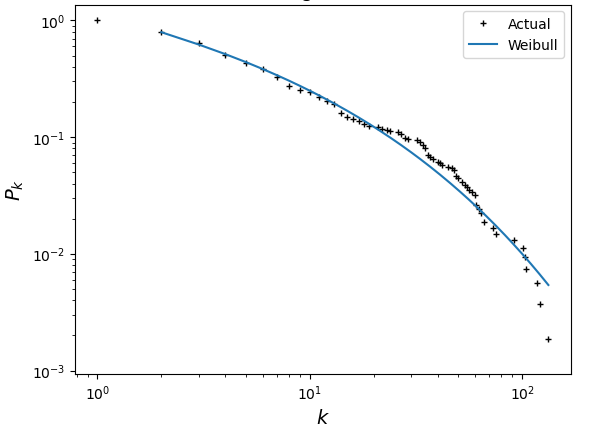}
        \caption{\textit{Mah\={a}bh\={a}rata} ($k_{min}=2$)}
    \end{subfigure}
    \hfill
    \begin{subfigure}[b]{0.32\textwidth}
        \centering
        \includegraphics[width=\textwidth]{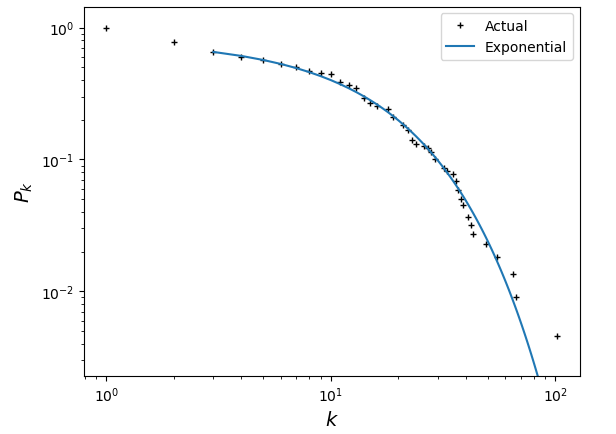}
        \caption{\textit{R\={a}m\={a}ya\d{n}a} ($k_{min}=3$)}
    \end{subfigure}
    \hfill
    \begin{subfigure}[b]{0.32\textwidth}
        \centering
        \includegraphics[width=\textwidth]{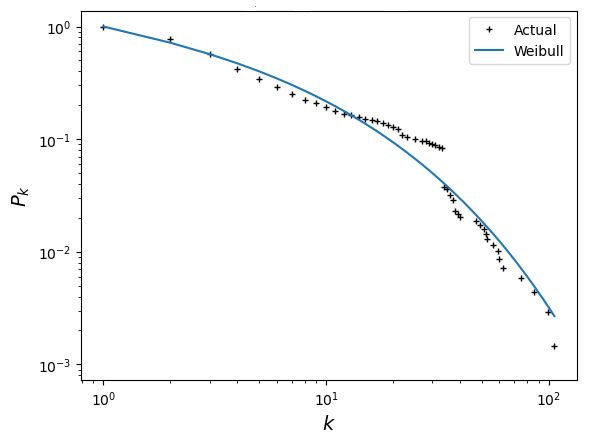}
        \caption{\textit{Iliad} ($k_{min}=1$)}
    \end{subfigure}

    \caption{Degree distributions for the three main networks (zero nodes excluded), fitted using maximum likelihood estimates.}
    \label{fig:distributions}
\end{figure*}

The \textit{degree} $k_i$ of an individual $i$ represents the number of edges linking the corresponding vertex to other vertices in the network. We then denote by $p(k)$ the probability that a chosen vertex has degree $k$. In many non-random networks it has been shown that their degree distributions are right-skewed. Indeed, it was argued by Barab\'{a}si and Albert \cite{BarabasiAlbert1999} that these distributions often follow a power law $p(k)\sim k^{-\gamma}$ for a positive constant $\gamma$ (typically in the range $2\leq \gamma\leq 3$). Since such distributions often exhibit noisy tails, we often look to the complementary cumulative distribution function so that $P(k)\sim k^{1-\gamma}$, \cite{AlbertBarabasi2002}. While popular at the time, nowadays power laws are seen as less likely. Even early on, Amaral et al. argued in \cite{Amaral2000} against the notion that social networks typically follow a power law. A significant contribution to this area was in \cite{ClausetShaliziNewman2009} in which Clauset, Shalizi and Newman raised concerns over many popular methods used to detect power laws, arguing a lot of power laws are in fact lognormal. More recently Broido and Clauset argue that power laws are rare, \cite{BroidoClauset2019}. Consequently, it is now less clear what degree distributions social networks have. We view it as likely that social networks tend to have right-skewed degree distributions and for our purposes this viewpoint is sufficient\footnote{In \cite{BroidoClauset2019}, the authors argue social networks are `at best weakly scale free'.}.

In \cite{BroidoClauset2019}, Broido and Clauset looked for for power laws in the tails of distributions. Such an approach is not without merit and indeed we argue it sensible to look at degrees greater than $k_{min}=1$. For if $k_{min}=1$, normalisation would require $\gamma<2$ and an expected mean degree which diverges. Since this is not what we typically see (average degrees tend to be relatively small), it is reasonable to use $k_{min}=2$. Nevertheless, the arguments of \cite{BroidoClauset2019} often only require a power-law to be a good fit to the 50 highest-degree nodes. This is clearly unsatisfactory in many ways and so we adopt the strategies employed in \cite{MannionMacCarron2023}. Building upon the method of maximum likelihood used in \cite{ClausetShaliziNewman2009}, Mannion and MacCarron provided a robust approach to identify the best fit for the distribution. Their focus went beyond power laws and included other distributions such as exponential ($p_k\sim e^{-k/\kappa}$), Weibull ($p_k\sim(k/\kappa)^{\beta-1}e^{(k/\kappa)^\beta}$), and lognormal ($p_k\sim\frac{1}{k}e^{-\frac{(\ln(k)-\mu)^2}{2\sigma^2}}$) distributions. Here, $\kappa$ is the rate parameter, while $\beta$ is the shape parameter. Note that when $\beta=1$, the Weibull distribution simplifies to the exponential distribution. When $\beta=2$, it is closely related to a normal distribution. Finally, $\mu$ is the mean and $\sigma^2$ the variance.

In \figref{fig:distributions}, we see the result of applying the methods of Mannion and MacCarron. In particular, we see that the \textit{Mah\={a}bh\={a}rata} follows a Weibull distribution whereas the \textit{R\={a}m\={a}ya\d{n}a} follows an exponential distribution. Note also that with a value of $\beta=0.37$, the \textit{Mah\={a}bh\={a}rata} is not particularly close to being exponential. As such, this represents a significant structural difference between the two networks. Furthermore, we see that there is a much slower rate of decay for the \textit{Mah\={a}bh\={a}rata}, indicating a wider spread of degrees.

To interpret this in the context of the narrative, we next turn our attention to clustering of nodes. The maximum possible pairs of neighbours of vertex $i$ is $\frac{k_i(k_i-1)}{2}$. If a pair of neighbours of $i$ are connected, then this forms a triangle with $i$ as one of the vertices. As such, we define the local clustering coefficient (see \cite{WattsStrogatz1998}) of $i$ to be

\[
C_i=\frac{2|C_3(i)|}{k_i(k_i-1)},
\]

\noindent where $C_3(i)$ is the set of triangles in which $i$ is a vertex. We can also think of $|C_3(i)|=n_i$ as the number of edges linking the $k_i$ neighbours of $i$. In a social network, $C_i$ measures the proportion of an individual's connections who are mutually acquainted. The \textit{mean clustering coefficient} $C$ is then defined to be the average of $C_i$ over all $N$ vertices.

If we take a random graph of the same size and average degree as our network, then we can compute the respective average path length $l_{rand}$ and average clustering coefficient $C_{rand}$ and compare. We then say a network is small world (see \cite{WattsStrogatz1998}) if $l\approx l_{rand}$ and $C>>C_{rand}$. Many complex networks share this property and so this is perhaps a poor distinguishing tool. Nevertheless, as we shall see, this does offer another insight into structural differences between the Indian texts.

\begin{table}[H]
\caption{Small world check.}
\label{tab:small_world}
\centering
    \begin{tabular}{c|ccccc}
    \toprule
    \textbf{Saga} & $l$ & $l_{rand}$ & $C$ & $C_{rand}$ \\
    \midrule
    \textit{Mah\={a}bh\={a}rata}  & 3.5586 & 2.9818 & 0.5090 & 0.0183 \\
    \textit{R\={a}m\={a}ya\d{n}a} & 2.6601 &  2.6225 & 0.5315 & 0.0407 \\
    \textit{Iliad} & 3.4842 & 3.4341 & 0.4380 & 0.0112 \\
    \bottomrule
    \end{tabular}
    \tabletext{Note: The random values were obtained by averaging 50 random simulations.}
\end{table}

In \tabref{tab:small_world}, we see that the \textit{R\={a}m\={a}ya\d{n}a} is small world whereas the \textit{Mah\={a}bh\={a}rata} could be deemed to fail to be small world given the larger average path lengths. This could suggest strong local structure in which there is insufficient long-range connectivity. However, given the scale of the difference (average path length is only slightly larger than the random counterpart), we can still argue that the \textit{Mah\={a}bh\={a}rata} is broadly small world. There exist shortcuts but not quite as many as in a random graph. This means we have inter-community links which keep path lengths short, but these are more structured than random.

Sticking with clustering, there is a sense in which taking an average of averages is less than ideal. This definition also tends to be dominated by vertices with low degree since they have small denominators. As such, it can be argued that this measures provides a rather poor picture of the overall properties of a network. Consequently, an alternative measure is often used due to Newman called \textit{transitivity}:

\[
C_T = \frac{\text{number of closed paths of length two}}{\text{number of paths of length two}}.
\]

\noindent This offers a global quantity as opposed to the local quantity defined above. As it happens, transitivity can be estimated from the degree distribution and we refer to this as the configuration-model estimate, $C_n$. While this estimate works reasonably well for non-social networks, the clustering into communities (see below) means that it typically fails for social networks. For social networks, we typically see $C_T>C_n$, \cite{NewmanPark2003}. As we see in \tabref{tab:network_properties_I}, both Indian texts exhibit this property.

We conclude this point on high clustering by providing the formula for $C_n$, as given in \cite{NewmanPark2003}. First, we recall the first and second moments are given by $\langle k\rangle =\frac{1}{N}\sum^N_{i=1}k_i$ and $\langle k^2\rangle =\frac{1}{N}\sum^N_{i=1}k^2_i$, respectively. Then,

\[
C_n=\frac{1}{N}\frac{(\langle k^2\rangle-\langle k\rangle)^2}{\langle k\rangle ^3}.
\]

\begin{table*}[t]
\centering
\caption{Comparative statistics for the three texts in which we look separately at the full network and friendly subnetwork. We denote the number of nodes by $N$, average path length by $l$, average degree by $\langle k\rangle$, transitivity by $C_T$ and the null configuration model estimate by $C_n$. We also use $r_k$ to represent the degree assortativity coefficient. Finally, we denote the giant component by $G_c$.}

\label{tab:network_properties_I}

\begin{tabular}{l|ccccccc|ccccccc}
\toprule
& \multicolumn{5}{c}{\textbf{Full network}}
& \multicolumn{5}{r}{\textbf{Friendly network}} \\
\cmidrule(lr){2-8} \cmidrule(lr){9-13}
\textbf{Saga}
& $N$ & $l$ & $\langle k \rangle$ & $C_T$ & $C_n$ & $r_k$ & $G_c (\%)$
& $N (\%)$ & $l$ & $\langle k \rangle$ & $r_k$ & $G_c (\%)$\\
\midrule
\textit{Mah\={a}bh\={a}rata} & 537 & 3.5586 & 9.9255  & 0.4218 & 0.2919 & -0.0487 & 519 (96.65) & 
515 (95.90) & 3.6461 & 9.6777 & -0.0147 & 491 (95.34)\\
\textit{R\={a}m\={a}ya\d{n}a} & 247 & 2.6601 & 10.1700 & 0.5184 & 0.2877 & -0.1102 & 202 (81.78) & 
204 (82.59) & 2.7902 & 11.2941 & -0.0154 & 181 (88.73)\\
\textit{Iliad}  & 693 & 3.4842 & 7.7000 & 0.4511 & 0.1310 & -0.0836 & 686 (98.99) & 636 (91.77) & 3.7801 & 7.2893 & 0.0964 & 546 (85.85)\\
\bottomrule
\end{tabular}
\end{table*}

Moving on, let $e$ be an edge, and let $k_{e_1}$ and $k_{e_2}$ be the degrees of the two vertices at each end of $e$. Then the mean degree of vertices at the end of an edge over the $M$ edges is given by,

\[
\bar{k} = \frac{1}{2M}\sum_{e=1}^M(k_{e_1}+k_{e_2}).
\]

\noindent We then relate $\langle k\rangle$ and $\bar{k}$ via $\bar{k}=\frac{\langle k^2\rangle}{\langle k\rangle}$. The \textit{degree assortativity} for the $M$ edges of an undirected graph is then given by,

\[
r_k=\frac{1}{M}\sum^M_{e=1}\frac{(k_{e_1}-\bar{k})(k_{e_2}-\bar{k})}{\sigma^2},
\]

\noindent where,

\[
\sigma=\sqrt{\frac{1}{2M}\sum^M_{e=1}(k_{e_1}-\bar{k})^2+(k_{e_2}-\bar{k})^2}.
\]

\noindent Note that $r_k$ necessarily falls between $-1$ and $1$, \cite{Newman2002}. If $r_k<0$, we say that the network is \textit{disassortative} and broadly means high-degree nodes tend to be connected to low-degree nodes, while if $r_k>0$ then we say the network is \textit{assortative}. Assortativity suggests high-degree (resp. low-degree) nodes tend to connect to other high-degree (resp. low-degree) nodes. In \cite{NewmanPark2003}, the authors argue that disassortativity is the `natural state' for all networks, and so the presence of assortativity corresponds to additional structure. They argue the presence of communities (see below) provides such a structure in social networks. It has also been suggested (cf. \cite{MacCarronKenna2012}) that disassortativity may reflect the conflictual nature of stories. Often, we see a character introduced solely to be killed by a hero or villain to demonstrate either the hero's might or the villain's threat. In so doing, we have a very high-degree node connecting to a low-degree node and this provides the disassortativity. In \cite{MacCarronKenna2012}, the point is therefore made that only in the friendly subnetworks can disassortativity be confidently used to signal artificiality. Looking at \tabref{tab:network_properties_I}, we see that both friendly and full networks of the Indian texts are mildly disassortative. We see that the \textit{R\={a}m\={a}ya\d{n}a} full network is the most disassortative, while its friendly subnetwork is only mildly disassortative. In contrast, the \textit{Mah\={a}bh\={a}rata} is only mildly disassortative in both the full and friendly network (although the friendly network is even less disassortative). Given, the heavily abridged nature of the \textit{Mah\={a}bh\={a}rata} in particular, we suspect the minimal change here is predominantly due to minor characters being omitted from the translation since they likely added little to the overall narrative.

Since we have distinguished between friendly and hostile edges, we can examine the \textit{structural balance} of our networks. This is the tendency of the network to disfavour triads with an odd number of hostile edges. This is due to the notion of `the enemy of my enemy is my friend', and that of the tendency for individuals to `pick a side'. For the \textit{Mah\={a}bh\={a}rata}, 13.04\% of triads have one hostile edge, while in the \textit{R\={a}m\={a}ya\d{n}a}, just 0.99\% have one hostile edge. In the \textit{Mah\={a}bh\={a}rata}, 0.01\% of triads have all hostile edges (12 in total), while no triads in \textit{R\={a}m\={a}ya\d{n}a} have all hostile edges. This difference is perhaps reflecting the differing nature of the two texts. In the {R\={a}m\={a}ya\d{n}a}, there is a broadly clear divide between the `heroes' and the `villains'. While there is in-fighting (e.g. R\={a}va\d{n}a's brother Vibh\={i}\d{s}a\d{n}a joining R\={a}ma, or Sugr\={i}va's fight with his brother V\={a}lin), largely this divide is clear. In contrast, the \textit{Mah\={a}bh\={a}rata} depicts a civil war between two sets of cousins. As such, we might expect a lot of in-fighting and this may explain the relatively high percentage of triads with an odd number of hostile edges. It is also telling that of the twelve triads with all negative edges, Duryodhana is present in six of these, while Kar\d{n}a is in four of the remaining six. These two characters are particularly divisive with Duryodhana being the main antagonist who is almost unapologetically villainous, while Kar\d{n}a is his most loyal and devoted supporter who famously fights with Bh\={i}\d{s}ma (Duryodhana's first commander in the war), refusing to fight while Bh\={i}\d{s}ma lives. Beyond R\={a}va\d{n}a, there are no other characters who hold a similar role in the \textit{R\={a}m\={a}ya\d{n}a}, and  R\={a}va\d{n}a largely carries absolute loyalty from his followers (R\={a}k\d{s}asas) who focus their anger at R\={a}ma and his army.

\begin{figure}[t]
    \centering
    \includegraphics[width=0.4\textwidth]{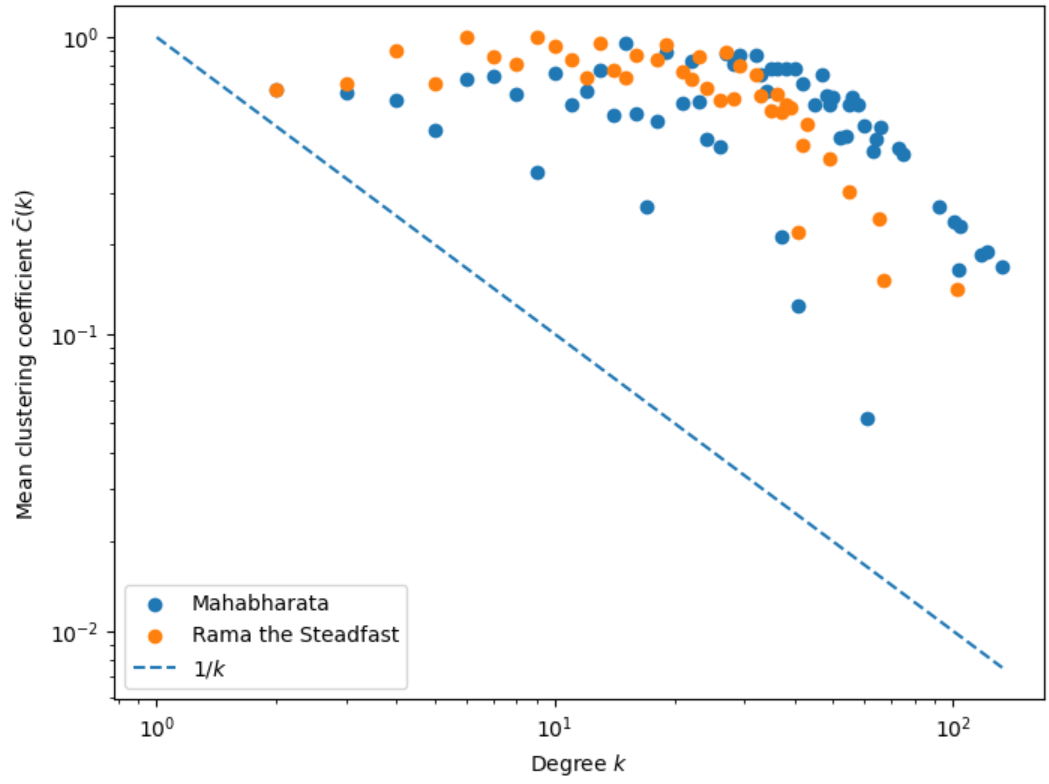}
    \caption{Mean clustering coefficient per degree for the two Indian texts. The power law $\frac{1}{k}$ is included as a dashed line to guide the eye.}
    \label{fig:hierarchical_clustering}
\end{figure}

It has been suggested that if the clustering coefficient $C_i$ decreases as a power of the degree $k_i$, then the network is hierarchical \cite{RavaszBarabasi2003}. In practice, this tends to not often be the case, and the notion of hierarchy in this context is not very well-defined. Nevertheless, a general decay does suggest that high degree vertices tend to have low clustering. Such nodes play a key role in connecting sub-communities since they have neighbours that are spread across otherwise weakly connected groups. As we see in \figref{fig:hierarchical_clustering}, neither network is following a power law. Indeed, for both texts, the distributions are remarkably stable in large parts of the distributions. This suggests far more cohesion than in a classic case of hierarchical clustering. While there is a broad pattern of lower-degree nodes having higher clustering in both texts, the situation is evidently far more complicated. What we see for both texts is a number of high degree nodes with high clustering, indicating highly interconnected groups of elites that likely overlap. This viewpoint would be consistent with court-like, dense interaction structures and that certainly matches the narratives. In Section \ref{sec:local_cohesion}, we shall examine this further and see that the situation is far more subtle than \figref{fig:hierarchical_clustering} suggests.

Next, we look for the presence of communities, though with the caveat that given the number of community detection algorithms available, we can find communities in most networks. As such, we employ three popular algorithms to detect communities. We stress that the purpose here is not to compare the outputs of these algorithms. Instead, we are attempting to see if there is any vague agreement between algorithms.

If we wish to understand connections between otherwise weakly connected groups, \textit{betweenness centrality} is a useful metric. This measures how many shortest paths (geodesics) pass through a given vertex, \cite{Freeman1977}. In a sense, it tells us how influential that vertex is in controlling the flow of information. If $\sigma(i,\,j)$ is the number of geodesics between vertices $i$ and $j$, and if the number of these which pass through a vertex $l$ is given by $\sigma_l(i,\,j)$, then the betweenness centrality of $l$ is given by,

\[
g_l=\frac{2}{(N-1)(N-2)}\sum_{i\neq j}\frac{\sigma_l(i,\,j)}{\sigma(i,\,j)}.
\]

\noindent The normalisation ensures that $g_l=1$ if all geodesics pass through vertex $l$.

Betweenness centrality can be used to detect community structures within a network (see \cite{GirvanNewman2002}), albeit using the similarly defined \textit{edge betweenness centrality}. This algorithm is commonly known as the Girvan-Newman algorithm and the core idea is that edges connecting communities will have a high edge betweenness. Alas, the Girvan-Newman algorithm is not always easy to interpret. As the algorithm proceeds, the network is successively broken down into $n$ components. The challenge is to optimise $n$. This can be done by visual inspection using a dendrogram, or by using modularity $Q$, \cite{NewmanGirvan2004}. To define $Q$, we first define $E$ to be an $n\times n$ matrix in which $E_{ij}$ is the proportion of all edges in the full network that link nodes in community $i$ to nodes in community $j$. If we denote $F_i=\sum_jE_{ij}$, the modularity is then defined by $Q=\sum_i(E_{ii}-F_i^2)$. If our network has just one community, $Q$ is close to zero. At the other extreme, where the network is partitioned into $n$ communities, each containing approximately $M/n$ edges, $Q\approx 1-\frac{1}{n}$. Thus, $Q$ is bounded by 1 for large $n$, though typically we see values between 0.3 and 0.7 for social networks containing community structure, \cite{NewmanGirvan2004}.

For the \textit{Mah\={a}bh\={a}rata}, the Girvan-Newman algorithm attained its maximum modularity at 117 communities with $Q=0.3520$, though it should be stressed that just twenty communities had more than 5 members and 63 communities had just a single member. This suggests a number of peripheral characters who have weak connections with the main network. In many ways this is unsurprising in a narrative network since minor characters are typically introduced and present only as a minor character in a single `episode'. The Girvan-Newman algorithm subsequently peels these characters off and places them into small communities.

As a first point of comparison, we use the Louvain community detection algorithm (see \cite{Blondel2008}) which maximises modularity. This produced just 18 communities for an optimum modularity of $Q=0.4614$ (unweighted) and $Q=0.3913$ (weighted). Again, most communities had few members with eight containing fewer than 4 members in the unweighted run of the Louvain algorithm, and ten containing fewer than 4 members in the weighted run. While such modularity maximisation algorithms are popular, there are several flaws worth noting, cf. \cite{FortunatoHric2016}, \cite{GhasemianHosseinmardiClauset2019}. One is that modularity maximisation tends to overfit significantly. Another is that, paradoxically, it can also lead to systematic underfitting. These can occur simultaneously so that areas of the network which is dominated by randomness can be found to contain communities, while other areas with a clear modular structure can have such communities obstructed. This naturally leads to several questions concerning the validity of any output. We therefore introduce a second point of comparison in the form of Infomap (see \cite{RosvallBergstrom2008}). This produced 57 communities for an optimum modularity of $Q=0.3894$ (unweighted), and 50 communities with $Q=0.2728$ (weighted). Again, most communities had few members with twenty containing fewer than 4 members in the unweighted run, while nineteen contained fewer than 4 members in the weighted run. We note that infomap also suffers from overfitting (see \cite{GhasemianHosseinmardiClauset2019}), and so view the Girvan-Newman algorithm as perhaps the most trustworthy of the three algorithms. Nevertheless, we argue that the combination of all three algorithms (Girvan-Newman, Louvain, Infomap) provide reasonable evidence for the existence of communities in the \textit{Mah\={a}bh\={a}rata}.

For the \textit{R\={a}m\={a}ya\d{n}a}, we have a similar picture. Optimising the Girvan-Newman algorithm yields 55 communities with a modularity score of $Q=0.4901$. Again, the majority of communities contain very few members with all but six containing fewer than five vertices (and the majority having just a single vertex). When using the Louvain algorithm, we obtain 16 communities with $Q=0.5433$ (unweighted), and 15 communities with $Q=0.4689$ (weighted). Using Infomap, we have 27 communities with $Q=0.5211$ (unweighted), and 26 communities with $Q=0.4616$ (weighted). As such, the network again appears to exhibit meaningful community structure, albeit alongside a large population of weakly integrated peripheral characters. The consistently larger modularity scores possibly indicate the \textit{R\={a}m\={a}ya\d{n}a} may possess somewhat stronger mesoscopic community structure than the \textit{Mah\={a}bh\={a}rata}. Care must be taken however, when comparing modularity scores in this way since modularity is sensitive to both network structure and the particular optimisation procedure employed. Nonetheless, the results tentatively suggest that interactions in the \textit{R\={a}m\={a}ya\d{n}a} may be somewhat more strongly concentrated within communities rather as opposed to distributed across them.

\begin{figure*}[t]
    \centering

    \includegraphics[scale=0.3]{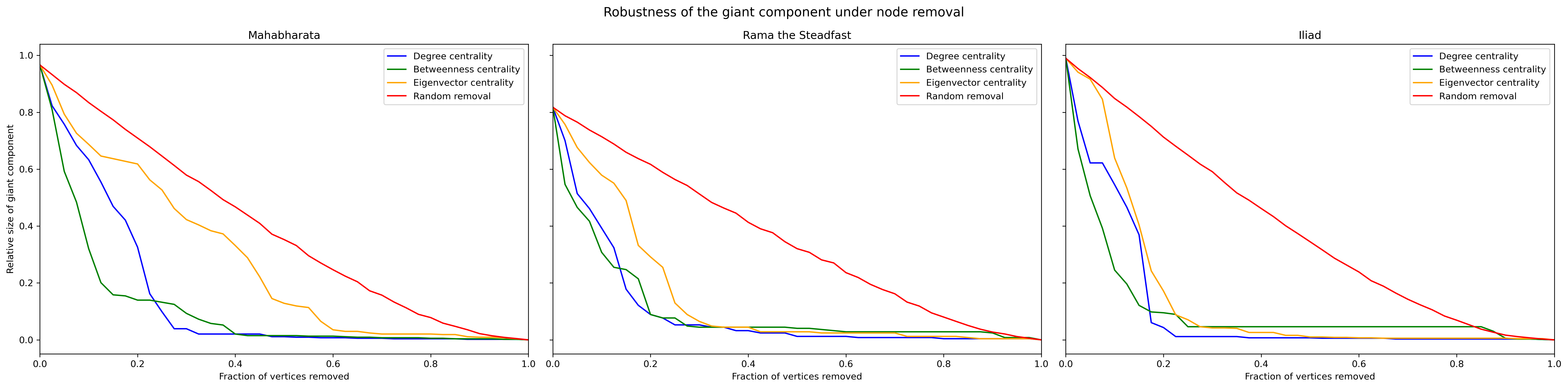}

    \caption{The relative size of the giant component as a fraction of the whole, plotted against the nodes removed as a fraction.}
    \label{fig:robustness}
\end{figure*}

To conclude this section, we discuss another use of betweenness centrality, namely that of \textit{robustness}. Many networks are found to be robust to random removal of vertices, but fragile to targeted removal (cf. \cite{AlbertBarabasi2002}, \cite{Newman2003}). To test this, we remove characters based upon one of three centrality measures, starting with the highest: betweenness, degree and eigenvector. We compare by also removing vertices randomly\footnote{We report the average effects of 100 realisations of random attack.}. After each removal, we measure the relative size of the giant component. What we observe in \figref{fig:robustness} is that both Indian networks are robust to random removal but fragile to attack based upon betweenness or degree centrality. This is much like other social networks. However, we note that the \textit{Mah\={a}bh\={a}rata} is more robust to removal based upon eigenvector centrality. To unpack this, we recall that eigenvector centrality captures the idea that a vertex is more central if it is connected to other important nodes. We posit that those vertices with high eigenvector centralities likely sit inside dense elite cores which are richly interconnected. As such, removing one is unlikely to break apart that cluster of the network since there are many other alternative paths available. Consequently, the network is more robust to removal based upon eigenvector centrality. This suggests another difference between the two Indian texts. In the \textit{Mah\={a}bh\={a}rata}, with its narrative based upon a highly interconnected set of cousins waging war upon each other, we see major characters interacting heavily with one another. Prestige is distributed across interconnected central figures and this makes high prestige characters relatively structurally redundant since they are not key information brokers. If we compare this with the \textit{R\={a}m\={a}ya\d{n}a}, there is perhaps a greater divide between the central characters; namely, we have R\={a}ma and his followers on one side, and R\={a}va\d{n}a and his followers on the other, with relatively fewer interconnections between these groups (until the war begins towards the end of the text). This means those characters who do bridge the two sides are far more structurally important.

\begin{table*}[t]
\caption{Top ten individual node removals ranked by reduction in giant component size after deleting a single node. Eigenvector centrality was computed on the original network before removal. We denote by $GC_0$ and $GC_1$ the size of the giant component pre- and post-removal, respectively.}
\label{tab:eigenvector_gc_attack}
\centering

\begin{tabular}{l|rccccc|l|rccccc}
\toprule
\multicolumn{7}{c|}{\textbf{\textit{Mah\={a}bh\={a}rata}}}
& \multicolumn{7}{c}{\textbf{\textit{R\={a}m\={a}ya\d{n}a}}} \\
\cmidrule(lr){1-7} \cmidrule(lr){8-14}

\textbf{Character} &
Rank &
Eig. Centrality &
GC$_0$ &
GC$_1$ &
$\Delta$GC &
$\Delta$GC (\%)
&
\textbf{Character} &
Rank &
Eig. Centrality &
GC$_0$ &
GC$_1$ &
$\Delta$GC &
$\Delta$GC (\%) \\
\midrule

Indra & 109 & 0.0256 & 519 & 502 & 17 & 3.28
&
R\={a}va\d{n}a & 33 & 0.0879 & 202 & 192 & 10 & 4.95 \\

Arjuna & 1 & 0.2015 & 519 & 509 & 10 & 1.93
&
Indra & 64 & 0.0349 & 202 & 193 & 9 & 4.46 \\

K\d{r}\d{s}\d{n}a & 15 & 0.1437 & 519 & 510 & 9 & 1.73
&
Vibh\={i}\d{s}a\d{n}a & 13 & 0.1790 & 202 & 196 & 6 & 2.97 \\

Vi\d{s}\d{n}u & 298 & 0.0010 & 519 & 510 & 9 & 1.73
&
Nala & 15 & 0.1726 & 202 & 197 & 5 & 2.48 \\

Brahm\={a} & 236 & 0.0032 & 519 & 511 & 8 & 1.54
&
Dundubhi & 176 & 0.0025 & 202 & 197 & 5 & 2.48 \\

Vasu Uparicara & 308 & 0.0008 & 519 & 512 & 7 & 1.35
&
R\={a}ma & 1 & 0.2424 & 202 & 198 & 4 & 1.98 \\

Cyavana & 150 & 0.0091 & 519 & 513 & 6 & 1.16
&
N\={i}la & 6 & 0.1898 & 202 & 199 & 3 & 1.49 \\

Yudhi\d{s}\d{t}hira & 3 & 0.1860 & 519 & 514 & 5 & 0.96
&
S\={i}ta & 39 & 0.0642 & 202 & 199 & 3 & 1.49 \\

Drupada & 26 & 0.1255 & 519 & 514 & 5 & 0.96
&
Himav\={a}n & 198 & 0.0001 & 202 & 199 & 3 & 1.49 \\

Dharma & 148 & 0.0099 & 519 & 514 & 5 & 0.96
&
Hanum\={a}n & 3 & 0.2004 & 202 & 200 & 2 & 0.99 \\

\bottomrule
\end{tabular}
\end{table*}

To test this, we removed a vertex and examined the change in giant component size. We then did this for all vertices and printed the top ten in terms of giant component reduction (\tabref{tab:eigenvector_gc_attack}). What is immediately obvious is that in both texts, the character who does most damage to the giant component does not have to be the character with the highest eigenvector centrality. Focusing on the \textit{Mah\={a}bh\={a}rata} first, we see that the top five characters in terms of damage to the giant component are Indra, Arjuna, K\d{r}\d{s}\d{n}a, Vi\d{s}\d{n}u and Brahm\={a}. Only two of those have an eigenvector centrality above 0.1 and Vi\d{s}\d{n}u and Brahm\={a} have eigenvector centralities of 0.00010 and 0.00032, respectively. This suggests there exists a distinction between core influence (captured by eigenvector centrality) and global connectivity (captured by the giant component). One possible explanation is that highly eigenvector-central characters occupy densely interconnected elite structures. While highly central characters might be influential, they are not irreplaceable bridges within the network. Furthermore, with several characters in the top 10 with very low eigenvector centrality (Indra ($1^{st}$), Vi\d{s}\d{n}u ($4^{th}$), Brahm\={a} ($5^{th}$), Vasu Uparicara ($6^{th}$), Cyavana ($7^{th}$), Dharma ($10^{th}$)), this suggests that some structurally important connectors occupy peripheral positions with respect to eigenvector centrality.

Turning to the \textit{R\={a}m\={a}ya\d{n}a}, the first observation is that characters in the top 10 have a larger impact on the giant component. The top two characters both result in a percentage drop of greater than 4\%. If we look at the top five then these all produce greater percentage drops than the top 10 for \textit{Mah\={a}bh\={a}rata}, except for Indra (who would make it to third in this list). We therefore cautiously claim that there are characters in the \textit{R\={a}m\={a}ya\d{n}a} who are more structurally important connectors, though the reader should not forget the difference in sizes of these two networks. Beyond this, the difference is more subtle. The \textit{R\={a}m\={a}ya\d{n}a} top-impact characters appear somewhat more strongly concentrated among moderate and highly eigenvector-central nodes, whereas the \textit{Mah\={a}bh\={a}rata} contains several highly disruptive nodes with comparatively small eigenvector centrality values. What this suggests is that in the \textit{Mah\={a}bh\={a}rata}, there are more highly disruptive nodes which have very small eigenvector centralities, i.e. in the \textit{Mah\={a}bh\={a}rata}, connective importance appears less tightly coupled to elite embeddedness than in the \textit{R\={a}m\={a}ya\d{n}a}.

\section{Local Cohesion}
\label{sec:local_cohesion}

In Section \ref{sec:universal_properties}, we saw two interesting structural differences between the Indian texts. When discussing potential hierarchical clustering, we saw a number of high degree nodes with high clustering, indicating highly interconnecting groups of elites which overlap. This is precisely the structure where we expect eigenvector-central nodes to be structurally replaceable, since removing one node does little because alternative paths exist. This was supported by our investigation into robustness which saw in particular that the \textit{Mah\={a}bh\={a}rata} was robust to the removal of nodes based on eigenvector centrality. To examine this further, look first to \tabref{tab:clustering_coefficient}.

\begin{table}[H]
\caption{Top 10 Characters by Clustering Coefficient.}
\label{tab:clustering_coefficient}
    \begin{tabular}{c|ccc}
    \toprule
    \textbf{Rank} & \textit{Mah\={a}bh\={a}rata} & \textit{R\={a}m\={a}ya\d{n}a} \\
    \midrule
    1 & Acala (1.0000) & Agastya (1.0000) \\
    2 & Adhiratha (1.0000) & Airavata (1.0000) \\
    3 & Adrika (1.0000) & Akampana (1.0000) \\
    4 & Adrika (Wife of Yayati) (1.0000) & Atri (1.0000) \\
    5 & Ahalya (1.0000) & Aviddha (1.0000) \\
    6 & Ahuka (1.0000) & Bhasakarna (1.0000) \\
    7 & Akarsa (1.0000) & Devantaka (1.0000) \\
    8 & Anadhrsti (1.0000) & Durdhara (1.0000) \\
    9 & Andhras (1.0000) & Durjaya (1.0000) \\
    10 & Angada (1.0000) & Durmukha (1.0000) \\
    \bottomrule
    \end{tabular}
\end{table}

The defining feature here is that all top ten characters exhibit maximal clustering coefficient. This is not especially uncommon and clustering coefficient is often seen as a poor discriminator for this reason. To address this, we looked instead at the percentage of those nodes with maximum clustering efficient. For the \textit{Mah\={a}bh\={a}rata}, this was 26.07\%, while for the \textit{R\={a}m\={a}ya\d{n}a} 28.74\% of all nodes had maximum clustering coefficient. These are substantial proportions, meaning that over one in four nodes exist in perfectly closed neighbourhoods. Note however that this does not mean a quarter of the graph is one giant clique. This is strictly a local statement. Nevertheless, it is a surprising proportion and suggests both texts demonstrate a tendency towards \textit{local cohesion} in which there are a large number of tightly closed neighbourhoods. This is consistent with a world in which many characters are introduced and embedded in small, highly interconnected groups.

\begin{table}[H]
\caption{Percentage of Characters with Maximal Clustering Coefficient.}
\label{tab:local_cohesion}
    \begin{tabular}{c|ccc}
    \toprule
     & \textit{Mah\={a}bh\={a}rata} & \textit{R\={a}m\={a}ya\d{n}a} \\
    \midrule
    Clustering $=1$ and degree $\geq 1$ & 26.07\% & 28.74\% \\
    Clustering $=1$ and degree $\geq 3$ & 16.01\% & 21.46\% \\
    Clustering $=1$ and degree $\geq 5$ & 8.94\% & 17.00\% \\
    \bottomrule
    \end{tabular}
\end{table}

However, this may also be reflective of a great many peripheral characters who have strong local cohesion, and therefore may have little to do with the more dominant characters. This is because lower-degree nodes need fulfil fewer properties to attain maximal clustering coefficient. For this reason, we filter by degree. As we see in \tabref{tab:local_cohesion}, this is to an extent the case for the \textit{Mah\={a}bh\={a}rata}. Once we exclude those nodes of degree $\leq 4$, less than 10\% of nodes have maximal clustering coefficient, compared to 17\% for the \textit{R\={a}m\={a}ya\d{n}a}. This may represent more self-contained local episodes of interaction in the \textit{R\={a}m\={a}ya\d{n}a} and could suggest a degree of artificiality, i.e. a sense in which the text is `too small world'. However, this is far from the only explanation for this behaviour. As such, we view this interpretation as speculative and nothing more.

\begin{figure*}[t]
    \centering
    
    \begin{subfigure}[b]{0.48\textwidth}
        \centering
        \includegraphics[height=6cm]{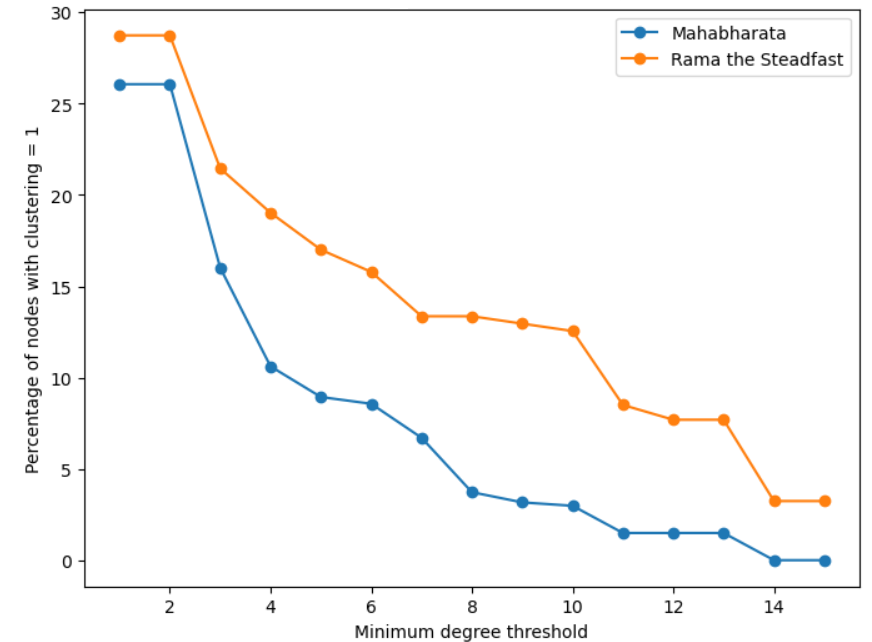}
        \caption{Nodes with clustering coefficient $1$ versus degree threshold.}
        \label{fig:local_cohesion_maha_rama}
    \end{subfigure}
    \hfill
    \begin{subfigure}[b]{0.48\textwidth}
        \centering
        \includegraphics[height=6cm]{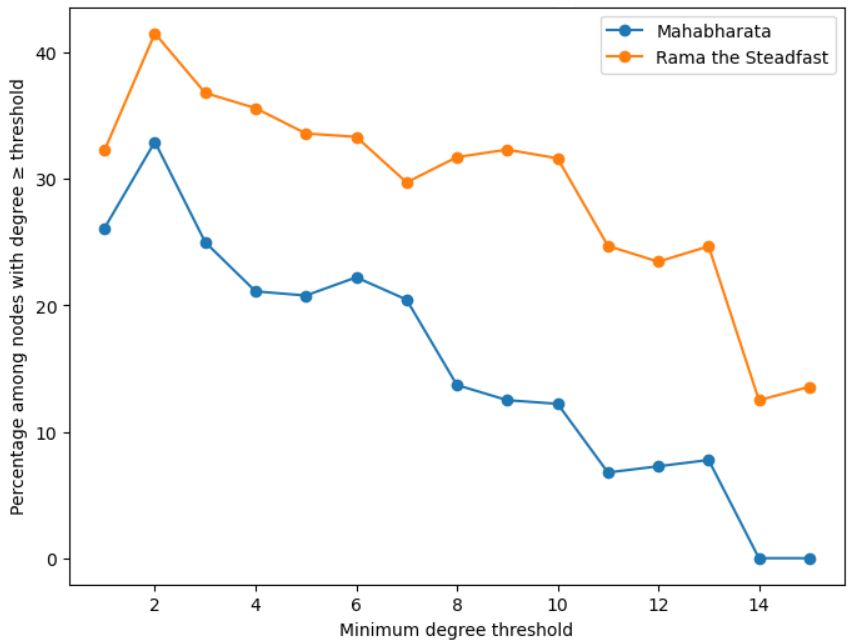}
        \caption{Share of perfectly clustered nodes among nodes with degree at least the specified threshold.}
        \label{fig:local_cohesion_maha_rama2}
    \end{subfigure}

    \caption{Persistence of perfect local cohesion among higher-degree nodes in the \textit{Mah\={a}bh\={a}rata} and \textit{R\={a}m\={a}ya\d{n}a}. In both figures, the \textit{R\={a}m\={a}ya\d{n}a} exhibits a consistently larger proportion of highly connected nodes with perfectly cohesive neighbourhoods.}
    \label{fig:local_cohesion}
\end{figure*}

To extend this, observe \figref{fig:local_cohesion}(a) in which we see a very clear distinction in the middle range (roughly degrees 4 to 13). The \textit{R\={a}m\={a}ya\d{n}a} clearly retains a greater percentage of nodes with maximal clustering coefficient in this region, suggesting more characters which are embedded in moderately large, fully interconnected groups. In contrast, the \textit{Mah\={a}bh\={a}rata} sees a much steeper decline and by $k\approx $ 8-10, very few high-degree nodes retain perfect local cohesion. Indeed, at high degree thresholds the percentage essentially collapses to zero. While it is not surprising that many high-degree \textit{Mah\={a}bh\={a}rata} characters act across partially disconnected groups, it does suggest that elite actors increasingly occupy positions which appear to connect partially distinct groups and may therefore adopt brokerage-oriented roles. This does not contradict our earlier findings that the text contains highly interconnecting groups of elites which overlap, and indeed may also help explain why the network is so fragile to betweenness centrality. Furthermore, since the \textit{R\={a}m\={a}ya\d{n}a} curve does not collapse quickly, this is consistent with high-degree characters in the \textit{R\={a}m\={a}ya\d{n}a} participating in dense cohesive structures, acting less as information brokers across communities. This provides a clear structural distinction from the \textit{Mah\={a}bh\={a}rata}.

To ensure we are not reading too much into \figref{fig:local_cohesion_maha_rama}, perhaps being misled by differing numbers of high-degree nodes in the Indian texts, we look to \figref{fig:local_cohesion_maha_rama2}. In this plot, we are asking among those nodes which are at least degree $n$, what fraction have perfectly closed neighbourhoods? As we see, the gap remains large throughout. This further supports our claims above and suggests that the difference between the two texts is not merely due to how many high-degree nodes they have. Instead, the difference appears to reflect a broader structural distinction between the two networks.

To conclude, the \textit{Mah\={a}bh\={a}rata} demonstrates a stronger tendency for cliques to break down as degree increases. This is consistent with the viewpoint the \textit{Mah\={a}bh\={a}rata} has a stronger tendency for high-degree characters to be information brokers acting in a distributed political structure. In contrast, in the \textit{R\={a}m\={a}ya\d{n}a} higher-degree nodes are substantially more likely to remain embedded in perfectly cohesive local neighbourhoods. In this sense, social cohesion persists even among prominent actors, indicating a narrative structure which is more unified and less politically fragmented. 

As a final remark in this section, we argue that figures such as those of \figref{fig:local_cohesion_maha_rama} may in some ways complement the standard hierarchical clustering plot. While hierarchical clustering plots are interesting in average clustering decay, our approach instead focuses on the sharper image of how persistent clique-like local organisation is in the network.

\section{Relationship with Western Texts}
\label{sec:relationship_with_western_texts}

Finally, we compare with the Western epic, the \textit{Iliad}. In particular, we shall focus on a comparison with the \textit{Mah\={a}bh\={a}rata} given the two texts are narratively similar. We preface what follows with a comment by Smith, \cite{SmithMaha} (pp. 1xv-1xvi): ``It may at first seem dangerously inappropriate to apply a Western genre-term such as `epic' to texts from a culture that recognizes no comparable genre, and that indeed assigns those texts to other genres...However, genres are not categories." As he goes on to argue, a genre ``is more like a bundle of typical characteristics, not all of which need apply in every case'' and as such, ``genre-terms are therefore available for making cross-cultural comparisons". We share Smith's viewpoint and therefore seek not to weigh in on whether the \textit{Mah\={a}bh\={a}rata} is suitable for the epic genre. Instead, we wish simply to make a broad structural comparison and to discuss the extent to which the two texts truly are similar. To do so, we shall roughly follow the arguments of Section \ref{sec:universal_properties}.

If we wish to start with some superficial statistics, we see from \tabref{tab:small_world} and \tabref{tab:network_properties_I} that the \textit{Iliad} has the greatest number of nodes of the three texts (693 against 537 and 247). Its average path length is 3.48 which is comparable to the \textit{Mah\={a}bh\={a}rata} (3.56) and larger than the \textit{R\={a}m\={a}ya\d{n}a} (2.66). Its giant component contains 98.99\% of nodes which again is far closer to the \textit{Mah\={a}bh\={a}rata} (96.65\%) than the \textit{R\={a}m\={a}ya\d{n}a} (81.78\%). The clustering coefficient is 0.4380 which is the smallest of the three texts (\textit{Mah\={a}bh\={a}rata} (0.5090), \textit{R\={a}m\={a}ya\d{n}a} (0.5315)), though all three are small world. Indeed, there is a slight difference between the \textit{Iliad} and the \textit{Mah\={a}bh\={a}rata}  in that the former, like the \textit{R\={a}m\={a}ya\d{n}a}, has $l\approx l_{rand}$. In contrast, the \textit{Mah\={a}bh\={a}rata} does have a somewhat larger average path length compared to $l_{rand}$, though the difference is not particularly large which is why we can still view the \textit{Mah\={a}bh\={a}rata} as small world. The \textit{Iliad} has a transitivity score of $0.4511$ against \textit{Mah\={a}bh\={a}rata}'s 0.4218 and \textit{R\={a}m\={a}ya\d{n}a}'s 0.5184. Interestingly, while the  \textit{Mah\={a}bh\={a}rata} exhibits higher average local clustering, the \textit{Iliad} has higher transitivity. This could be because characters in the \textit{Mah\={a}bh\={a}rata} inhabit more tightly knit local communities. These communities may be small and peripheral, and high-degree characters then act as bridges between them. In contrast, in the \textit{Iliad}, triangle formation is concentrated around the central cast, producing a denser core despite lower clustering coefficient for the average character. All three texts have high clustering coefficient.

\begin{figure}
    \centering
    \includegraphics[width=\linewidth]{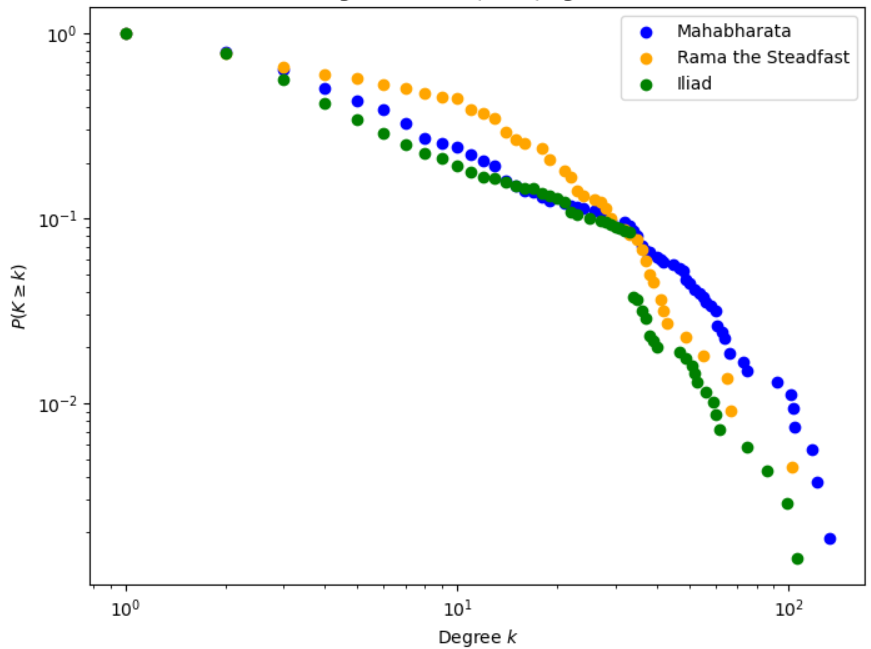}
    \caption{Degree CCDF: $P(K\geq k)$ against $k$.}
    \label{fig:deg_dist_all_three}
\end{figure}

To dig deeper we start with the degree distributions. In \figref{fig:distributions} (see also \figref{fig:deg_dist_all_three}), both \textit{Mah\={a}bh\={a}rata} and the \textit{Iliad} are best fit by a Weibull distribution (recall, \textit{R\={a}m\={a}ya\d{n}a}'s best fit is exponential). However, there are key differences. For example, for the \textit{Mah\={a}bh\={a}rata}, the shape parameter is below 1 (0.90) and so this indicates a stretched exponential tail which supports the interpretation that the network is relatively hub-heavy. High-degree characters are more common than under an exponential decay model and in this sense, the \textit{Mah\={a}bh\={a}rata} appears more hub-dominated than the \textit{R\={a}m\={a}ya\d{n}a}. In contrast, for the \textit{Iliad}, the shape parameter is above 1 (2.05) which suggests a compressed tail relative to an exponential, i.e. the tail is decaying faster. There are perhaps fewer very-high degree nodes than in the \textit{Mah\={a}bh\={a}rata}. It therefore appears more degree-homogeneous than the \textit{Mah\={a}bh\={a}rata} and has less extreme hub dominance. To summarise, on these fits, the \textit{Mah\={a}bh\={a}rata} has the strongest hub-dominated structure, \textit{R\={a}m\={a}ya\d{n}a} is closer to a standard exponential hierarchy, and the \textit{Iliad} is the least heavy-tailed of the three. In particular, \figref{fig:deg_dist_all_three} makes it easy to see that it is the tail in which the vast majority of differences take place. In the early half of the distribution, both the  \textit{Mah\={a}bh\={a}rata} and the \textit{Iliad} are remarkably similar. However, in the latter half we see the the \textit{R\={a}m\={a}ya\d{n}a} and the \textit{Iliad} become more similar. We therefore conclude that a major structural distinction between the texts lies in the extent and prominence of high degree hubs.

One aspect of the network which differentiates the \textit{Iliad} from the two Indian texts is degree assortativity. All three texts are disassortative with the \textit{R\={a}m\={a}ya\d{n}a} the most disassortative of the three. However, upon looking to the friendly subnetwork which we recall is in some sense a more suitable network to discuss assortativity, we see that the \textit{Iliad} become mildly assortative. This indicates that the \textit{Iliad} is particularly affected by low-degree nodes interacting with high-degree nodes, likely characters introduced for the sole reason of fighting a hero. However, we argue this difference may be a product of the translations used for the Indian texts. For the \textit{Mah\={a}bh\={a}rata} in particular, we use a substantially shorted version. It is likely that during the shortening process, Smith cut many such low-degree nodes as they are unlikely to drive the narrative forward. This may possibly explain why the \textit{Mah\={a}bh\={a}rata} sees the smallest increase in assortativity when moving to the friendly subnetwork.

Following Section \ref{sec:universal_properties}, we next look to structural balance. Recall, the \textit{Mah\={a}bh\={a}rata} has 13.04\% of triads with one hostile edge and 0.01\% with three hostile edges. In the \textit{R\={a}m\={a}ya\d{n}a}, 0.99\% of triads have one hostile edge and none had three. The \textit{Iliad} falls somewhere in between with 4.33\% of triads having one hostile edge and no triads having three edges. The reasoning we gave in Section \ref{sec:universal_properties} is consistent here. In the \textit{Iliad}, for the most part there is a clear duality to the conflict. On the one hand, the Trojans. On the other, the Greeks. This mirrors the \textit{R\={a}m\={a}ya\d{n}a} in which we have R\={a}ma and his followers on one side, and R\={a}va\d{n}a and his followers on the other. In contrast, the \textit{Mah\={a}bh\={a}rata} is a lot more complicated since it involves a civil war between two sets of cousins, and features conflicting loyalties.

\begin{figure*}[t]
    \centering

    \begin{subfigure}[b]{0.45\textwidth}
        \centering
        \includegraphics[width=\textwidth]{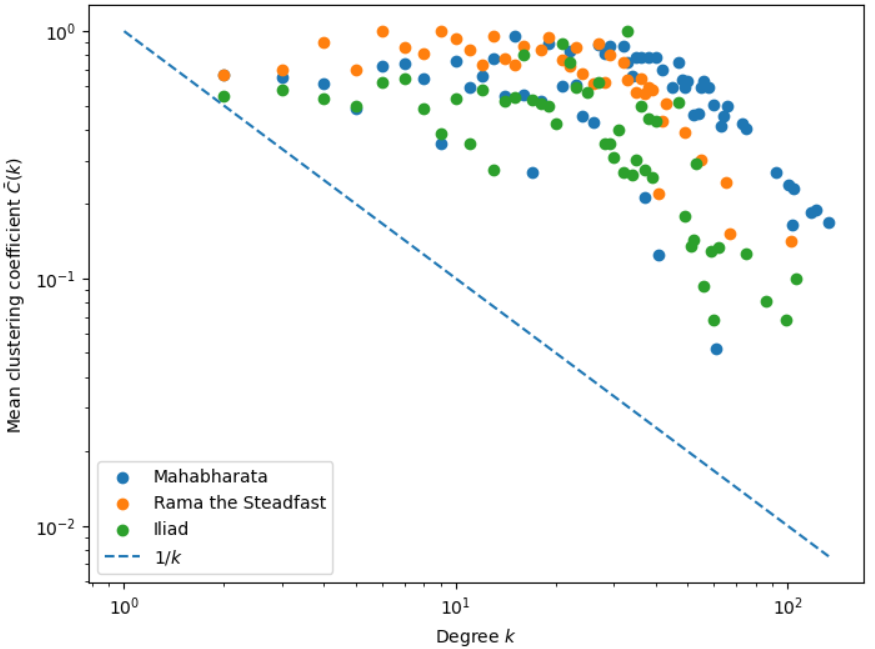}
        \caption{All three texts.}
        \label{fig:hier_all_three1}
    \end{subfigure}
    \hfill
    \begin{subfigure}[b]{0.45\textwidth}
        \centering
        \includegraphics[width=\textwidth]{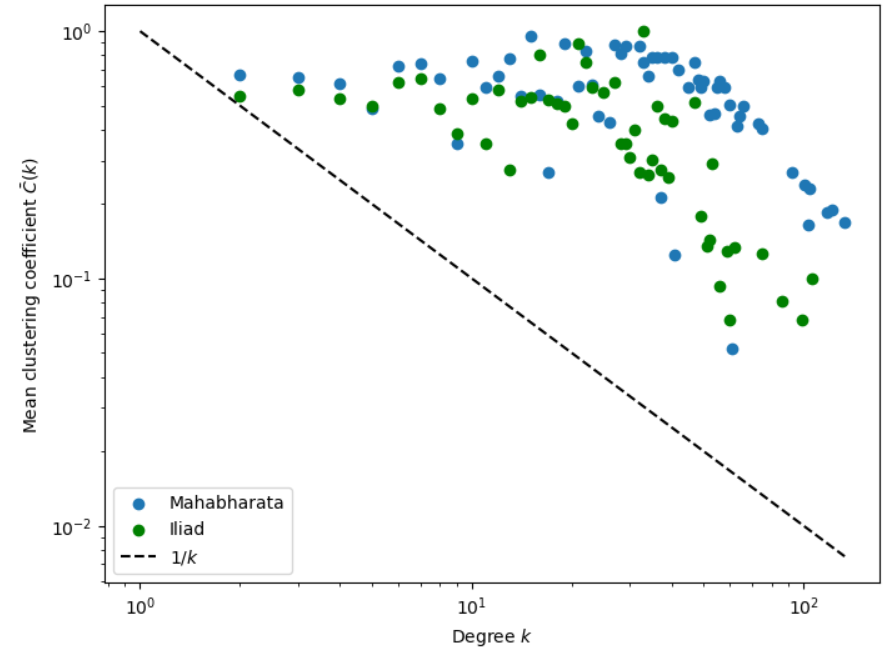}
        \caption{\textit{Mah\={a}bh\={a}rata} and the \textit{Iliad}.}
        \label{fig:hier_all_three2}
    \end{subfigure}

    \caption{Two plots of mean clustering coefficient against degree. As before, we include the line $\frac{1}{k}$ to guide the eye.}
\end{figure*}

Recall, we saw that the clustering coefficients of the two Indian texts do not follow a strict power law. Nevertheless, they do see a broad decay which supports high degree nodes tending to have lower clustering. As we have seen, the behaviour of high degree nodes provides a distinguishing feature between the three texts. In \figref{fig:hier_all_three1}, we view all three texts, while in \figref{fig:hier_all_three2} we isolate the \textit{Iliad} and the \textit{Mah\={a}bh\={a}rata}. The figures demonstrate a consistent picture in which the early stages of all three networks follow a broadly similar pattern. However, it is the behaviour for high-degree nodes which demonstrates the greatest differences. The \textit{Iliad} shows the steepest drop off implying high-degree nodes have lower clustering than the Indian texts (note, however there are exceptions). The \textit{Mah\={a}bh\={a}rata} has for the most part the slowest drop off rate, maintaining comparatively high clustering over a broader degree range, including among very high-degree vertices, and the \textit{R\={a}m\={a}ya\d{n}a} lies somewhere in between. This is consistent with our earlier findings that the \textit{Mah\={a}bh\={a}rata} contains more prominent high degree hubs. Combined with the earlier degree distribution analysis, this suggests the \textit{Iliad} exhibits the weakest tendency toward hub dominance of the three texts. This represents a significant structural difference in the organisation of high degree nodes between the Indian texts, and the \textit{Mah\={a}bh\={a}rata} in particular. Furthermore, as the \textit{Iliad's} high-degree nodes tend to lose clustering more rapidly than either of the Indian texts, this is consistent with our view that it is less locally cohesive than the \textit{R\={a}m\={a}ya\d{n}a} and more degree-homogeneous.

\begin{figure*}[t]
    \centering
    
    \begin{subfigure}[b]{0.48\textwidth}
        \centering
        \includegraphics[height=6cm]{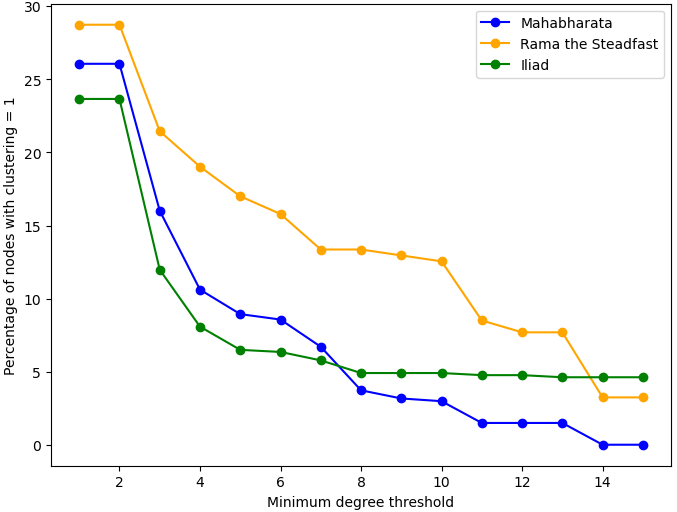}
        \caption{Nodes with clustering coefficient $1$ versus degree threshold.}
        \label{fig:local_cohesion_maha_rama_iliad1}
    \end{subfigure}
    \hfill
    \begin{subfigure}[b]{0.48\textwidth}
        \centering
        \includegraphics[height=6cm]{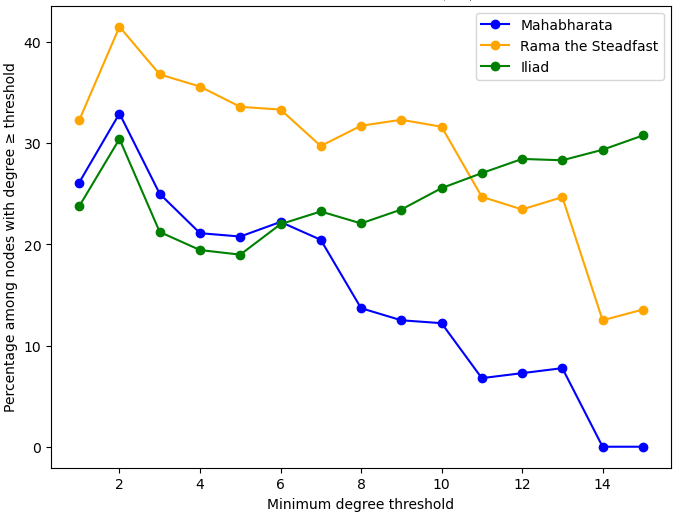}
        \caption{Share of perfectly clustered nodes among nodes with degree at least the specified threshold.}
        \label{fig:local_cohesion_maha_rama_iliad2}
    \end{subfigure}

    \caption{Persistence of perfect local cohesion among higher-degree nodes in the \textit{Mah\={a}bh\={a}rata}, \textit{R\={a}m\={a}ya\d{n}a} and \textit{Iliad}.}
    \label{fig:local_cohesion_maha_rama_iliad}
\end{figure*}

To examine local cohesion more we repeat what we did for the Indian texts. As a starting point, note 23.67\% of nodes in the \textit{Iliad} have maximal clustering compared to 26.07\% and 28.74\% for the \textit{Mah\={a}bh\={a}rata} and \textit{R\={a}m\={a}ya\d{n}a}, respectively. Once we only look for nodes of degree $\geq 3$, the percentage drops to 11.98\% for the \textit{Iliad}, compared to 16.01\% and 21.46\% for the \textit{Mah\={a}bh\={a}rata} and \textit{R\={a}m\={a}ya\d{n}a}, respectively. Once we look at nodes of degree $\geq 5$ the percentage further drops to 6.49\%, compared to 8.94\% and 17\% for the \textit{Mah\={a}bh\={a}rata} and \textit{R\={a}m\={a}ya\d{n}a}, respectively. Observe \figref{fig:local_cohesion_maha_rama_iliad1}, in which the \textit{Iliad} curve starts below both Indian texts but remains remarkably stable from $\approx degree =5$, eventually finishing above the curves of both Indian texts. This suggests it may be less globally clique-heavy than the \textit{R\={a}m\={a}ya\d{n}a} but also less dominated by large bridging hubs than the \textit{Mah\={a}bh\={a}rata}. This pattern is sharpened in \figref{fig:local_cohesion_maha_rama_iliad2} in which we see the conditional percentage for the \textit{Iliad} rise sharply after  $\approx degree =5$. This suggests the \textit{Iliad} contains a subset of high-degree characters whose neighbourhoods remain unusually cohesive despite relatively large degree. So, although the \textit{Iliad} is less hub-heavy overall than the \textit{Mah\={a}bh\={a}rata}, its important local elites may be embedded within highly interconnected military or kinship circles. One possible explanation for these differences is that the \textit{Mah\={a}bh\={a}rata} contains a broader population of high-degree nodes than either the \textit{R\={a}m\={a}ya\d{n}a} or the \textit{Iliad}. Connectivity is therefore distributed across a larger set of structurally important characters, which may contribute both to the network’s robustness under eigenvector-based attack and to the reduced prevalence of nodes with perfect local clustering. High-degree vertices in the \textit{Mah\={a}bh\={a}rata} still tend to exhibit substantial clustering, but because many act as connectors between partially distinct narrative communities, fewer retain clustering coefficient $=1$. In contrast, if the \textit{R\={a}m\={a}ya\d{n}a} and \textit{Iliad} contain fewer high-degree nodes overall, highly connected characters more frequently remain embedded within tightly interconnected local neighbourhoods, leading to a greater proportion of nodes with maximal clustering.

To test this hypothesis, we examined the proportion of total network degree accounted for by the highest-degree vertices in each text. Specifically, for each network, vertices were ranked by degree and the top 10\% of nodes were identified. We then computed the proportion of the network's total degree contained within this subset. Formally, if $V_{0.1}\subseteq V$ denotes the set consisting of the top 10\% of vertices ranked by degree, and $k_v$ denotes the degree of vertex $v$, then the degree concentration is computed as

\[
\frac{\sum_{v\in V_{0.1}} k_v}
{\sum_{v\in V} k_v}.
\]

This quantity measures the extent to which connectivity is concentrated among a relatively small elite subset of characters. Higher values indicate stronger hub dominance, whereas lower values indicate a more homogeneous degree structure. Interestingly, the results were somewhat mixed. The degree concentrations were 54.62\%, 41.08\% and 51.07\% for the \textit{Mah\={a}bh\={a}rata}, \textit{R\={a}m\={a}ya\d{n}a} and \textit{Iliad}, respectively. The \textit{Mah\={a}bh\={a}rata} exhibits the strongest degree concentration, with over half of all connectivity contained within the top 10\% of characters. However, the difference with the \textit{Iliad} is not as substantial as expected, with the \textit{Iliad} also exhibiting high degree concentration, albeit to a lesser extent. Perhaps a more refined viewpoint is to say the \textit{Iliad}'s major hubs are concentrated within tightly interconnected elite structures and function less as global integrative bridges than those of the \textit{Mah\={a}bh\={a}rata}, or at least not to the same extent. In contrast, the substantially lower concentration observed in the \textit{R\={a}m\={a}ya\d{n}a} (41.08\%) supports the viewpoint that connectivity is distributed more evenly, with its influential characters more locally embedded.

Finally, we consider communities within the \textit{Iliad}. With the Girvan-Newman algorithm we find that the \textit{Iliad} attains a best split with 70 communities and a modularity of 0.5903. Recall, the \textit{Mah\={a}bh\={a}rata} and \textit{R\={a}m\={a}ya\d{n}a} attained best split modularities of 0.3520 and 0.4901, respectively. As with the Indian texts, the majority of communities have fewer than 5 members. When using the Louvain algorithm, the best split produced 16 communities and a modularity of 0.6292 (unweighted), and 22 communities with modularity 0.6264 (weighted). As such, we again have higher modularity than the Indian texts: \textit{Mah\={a}bh\={a}rata} (0.4614 and 0.3913) and \textit{R\={a}m\={a}ya\d{n}a} (0.5433 and 0.4689). Finally, when using Infomap the \textit{Iliad} has 82 communities with modularity $0.5846$ (unweighted), and 92 communities with modularity $0.5929$ (weighted). For the \textit{Mah\={a}bh\={a}rata}, this algorithm produced 57 communities and modularity $0.3894$ (unweighted), and 50 communities and modularity $0.2728$, while the \textit{R\={a}m\={a}ya\d{n}a} had 27 communities with modularity $0.5210$ (unweighted), and 26 communities with modularity $0.4616$ (weighted). Again, we should not ignore the issues associated with Louvain and Infomap. However, coupled with the results of the Girvan-Newman algorithm, we do see that the \textit{Iliad} attained consistently higher modularity scores than either Indian text, possibly indicating a stronger community structure. Of course, comparing modularity in this way is notoriously sensitive to other factors. As such, we feel it safer to simply conclude that all three texts consistently exhibit varying degrees of community structure.

\section{Conclusion}
\label{sec:conclusion}

Recall Smith's words regarding the puzzling placement of the \textit{R\={a}m\={a}ya\d{n}a} and the \textit{Mah\={a}bh\={a}rata} into different genres. His explanation is that it likely has `more to do with the strong perception of V\={a}lm\={i}ki, author of the \textit{R\={a}m\={a}ya\d{n}a}, as the founder of Sanskrit literary tradition than with any sense that the two works are radically different in character.' Our analysis here broadly agrees with this view and we found no overwhelming structural evidence to support a historical/poetic divide. Both texts share some properties commonly associated with real-world social networks although both are disassortative. Nevertheless, we still found several key differences. For instance, the \textit{Mah\={a}bh\={a}rata}'s robustness to removal based upon eigenvector centrality, indicating greater interconnection among central figures, likely reflecting the narrative setting of two pairs of cousins waging war upon each other. When investigating local cohesion, we again saw differences in which the \textit{Mah\={a}bh\={a}rata} demonstrated elite actors becoming increasingly brokerage oriented. In contrast, the \textit{R\={a}m\={a}ya\d{n}a} seems to be much more locally cohesive, possibly due to more self-contained local episodes of interaction. As we have noted, this could represent a sense in which the \textit{R\={a}m\={a}ya\d{n}a} is `too small world'. However, we do not see this as sufficient evidence of the supposed poetic/historical divide. While it is suggestive, the difference between the two could simply reflect different narrative choices. In the \textit{Mah\={a}bh\={a}rata}, there is a civil war and so it is perhaps unsurprising that there is less local cohesion than in the \textit{R\={a}m\={a}ya\d{n}a} which has a well established duality between the protagonist's group and the antagonist's. That this difference should be due to the poetic/historical divide is, we believe, too strong a claim.

Indeed, we see far more similarities than differences between the two texts. Both are small world with a high clustering coefficient and low numbers of triangles with odd numbers of hostile edges, although the \textit{R\={a}m\={a}ya\d{n}a} does have substantially fewer such triads. Both also display community structure and are robust to random attack, though fragile to more targeted attack (especially attack based upon betweenness centrality). For this reason, we opt to side more with Smith's viewpoint. While differences certainly exist between the two epics, these are not especially radical.

The other main point of comparison is with the \textit{Iliad}. As discussed in Section \ref{sec:relationship_with_western_texts}, there are several commonalities between the \textit{Iliad} and the Indian texts. All three networks exhibit small-world structure together with high clustering coefficients. None display clustering behaviour consistent with a power-law decay, and all exhibit varying degrees of community structure. However, we also observed several significant differences, particularly in the manner in which elite hubs interact with the wider network. This appears to be one of the principal structural distinctions between the \textit{Iliad} and the \textit{Mah\={a}bh\={a}rata} in particular. In the \textit{Mah\={a}bh\={a}rata}, highly inter-connected elite character groups play important roles as integrative bridges between partially distinct narrative communities. One consequence of this structure is that the network is substantially more robust to attack based on eigenvector centrality. More generally, the \textit{Mah\={a}bh\={a}rata} appears to exhibit a more intricate pattern of interactions among its elite characters than either of the other texts, perhaps reflecting the civil-war nature of its narrative and the overlapping affiliations this produces. By contrast, the \textit{Iliad} and the \textit{R\={a}m\={a}ya\d{n}a} possess relatively well-defined opposing sides, which may contribute to the more locally cohesive and structurally predictable behaviour of their elite hubs. Nevertheless, from the perspective of network structure, we find no evidence to deny the suitability of the label `epic' for the Indian texts. In particular, the \textit{Mah\={a}bh\={a}rata} and the \textit{Iliad} share a number of important structural characteristics alongside their well-known narrative similarities.

As a final comment, we should address the issues with the translations chosen, both of which involve a heavy amount of abridgement. It is an open problem to what extent different translations affect structural properties. Once different versions and/or abridgements are entered into the mix, the questions of representativeness only increase. As such, we see a future direction involving a thorough analysis of networks relating to various translations and, most importantly, fuller versions of the text. We should stress the monumental effort such an undertaking would represent, owing to the continued addition and modification of the texts upon copying. For example, while the oldest \textit{R\={a}m\={a}ya\d{n}a} manuscript dates from 1020CE, copies were formed as late as the Seventeenth and Eighteenth Centuries and they bear evidence of R\={a}m\={a}'s increasing veneration. A tantalising possibility may therefore be to explore the structural differences between the two main recensions which J. and M. Brockington describe as roughly corresponding to a Southern group and a Northern group. While the Southern group is traditionally regarded as remaining closer to the \textit{R\={a}m\={a}ya\d{n}a}'s origins, the degree of cross-fertilization within and between recensions and sub-recensions is far from insignificant.

\section{Appendix: Pronouncing Sanskrit}
\label{app:pronouncing_sanskrit}
 
When written in Roman script, Sanskrit words require the use of accented characters. While not strictly necessary for the above, it may be beneficial to understand some basic rules for such accented characters, given the oral history of these texts. We closely follow the relevant sections of \cite{BrockingtonRama} and \cite{SmithMaha}, providing only a brief overview of what we have found useful when reading the aforementioned texts.

For vowels, a macron (e.g. \={a}, \={i}, \={u}) indicates a long vowel, thus:
\begin{itemize}
    \item `a' is similar to `u' in English `sun', while `\={a}' is similar to the first `a' in `saga';
    \item `i' is similar to the `i' in `sit', while `\={i}' is like the `ee' in `seem';
    \item `u' is like the `oo' in `soot', while `\={u}' is like the `oo` in `soon';
    \item `e' is like the vowel of `say';
    \item `o' is like the vowel of `so';
    \item The dipthongs `ai' and `au' are like the vowels of `sight' and `sound', respectively;
    \item In Sanskrit, `\d{r}' is a vowel which in North India is pronounced as `ri', while in South India is pronounced as `ru'.
\end{itemize}
Note that the majority of names end in a vowel. Typically, male names end in a short vowel (-a, -i or occasionally -u), while female names end in a long vowel (-\={a}, -\={i}, -\={u}). Importantly, the length of a vowel is not related to the stress in pronunciation. All syllables are given equal stress.

For consonants, we have the following:
\begin{itemize}
    \item Dots below the letters `t,\,d,\,n,\,s' indicate retroflexion. In India, these are pronounced with the tongue curled upwards. For `s', this makes the sound similar to `sh' in `ship';
    \item An acute accent `\'{s}' on `s' represents a second, slightly different (non-retroflex) `sh' sound,
\end{itemize}
While there are other accents, including dots above letters, these can be largely ignored by the general reader.

To put the above into practice, K\d{r}\d{s}\d{n}a would be pronounced `K\d{r}\d{s}\d{n}a' if using North Indian pronunciation, or `Krushna' if using South Indian pronunciation.

For unaccented characters, we have the following:
\begin{itemize}
    \item `c' is pronounced like `ch' in `chip';
    \item `h' is used in combination with many other consonants to indicate aspiration, e.g. `th' is not pronounced as in `the' but instead is pronounced more similarly to `t' or as in `goatherd', or `ph' is pronounced as in `uphill', not `cough'.
\end{itemize}

Finally, the first and final syllables of a name can be modified in various ways to represent the son or daughter of a character. For example, a son of Da\'{s}aratha is D\={a}\'{s}arathi, the son of R\={a}va\d{n}a is R\={a}va\d{n}i, and so on. A female example would be the daughter of Janaka being J\={a}nak\={i}.

\section{Acknowledgements}

The author would like to warmly thank P\'{a}draig MacCarron who kindly provided the \textit{Iliad} and \textit{Mah\={a}bh\={a}rata} datasets. The author would also like to remember Ralph Kenna who first introduced me to this area of research.

%----------------------------------------------------------

\printbibliography

%----------------------------------------------------------

\end{document}